\documentclass[11pt, preprint]{aastex}

\def\gs{\mathrel{\raise0.35ex\hbox{$\scriptstyle >$}\kern-0.6em
\lower0.40ex\hbox{{$\scriptstyle \sim$}}}}
\def\ls{\mathrel{\raise0.35ex\hbox{$\scriptstyle <$}\kern-0.6em
\lower0.40ex\hbox{{$\scriptstyle \sim$}}}}

\shorttitle{$\rm H_2$ in numerical simulations}
\shortauthors{Pelupessy, Papadopoulos, \& van der Werf}

\begin{document}

\title{Incorporating the molecular gas phase  in  galaxy-size numerical
simulations: first applications in dwarf galaxies}

\author{Inti Pelupessy}
\affil{Sterrewacht Leiden,  P. O. Box 9513,  2300 RA Leiden, The Netherlands,\\
Carnegie-Mellon University, Department of Physics, 5000 Forbes Avenue,
Pittsburgh PA,15213}
\email{pelupes@andrew.cmu.edu}

\author{Padeli \ P.\ Papadopoulos}
\affil{Institut f\"ur Astronomie, ETH Zurich, 8093 Z\"urich, Switzerland}
\email{papadop@phys.ethz.ch}

\and 

\author{Paul van der Werf}
\affil{Sterrewacht Leiden,  P. O. Box 9513,  2300 RA Leiden, The Netherlands}
\email{pvdwerf@strw.leidenuniv.nl}

\begin{abstract}

We present models of the  coupled evolution of the gaseous and stellar
content of galaxies using  a hybrid N-body/hydrodynamics code, a Jeans
mass  criterion  for the  onset  of  star  formation from  gas,  while
incorporating for the first time the  formation of H$_2$ out of HI gas
as part of such a model.  We  do so by formulating a subgrid model for
gas clouds that uses well-known cloud scaling relations and solves for
the  HI$\leftrightarrow$H$_2$ balance  set by  the H$_2$  formation on
dust grains and its FUV-induced photodissociation by the temporally and
spatially  varying  interstellar  radiation  field.   This  allows  the
seamless  tracking  of the  evolution  of  the  H$_2$ gas  phase,  its
precursor Cold  Neutral Medium (CNM) HI gas,  simultaneously with the 
star  formation.  An  important  advantage  of incorporating the 
molecular gas phase in numerical studies of galaxies is that the  set 
of observational constraints becomes  enlarged by the widespread 
availability  of H$_2$ maps  (via its tracer  molecule CO).
We then  apply our model  to the description  of the evolution  of the
gaseous  and stellar content  of a typical  dwarf galaxy. Apart from
their importance in galaxy evolution,  their small size allows  our 
simulations to  track  the thermal  and dynamic evolution of gas as 
dense as $\rm  n\sim 100\ cm ^{-3}$ and as cold as 
$\rm  T_{k}\sim 40\  K$,  where  most of  the  $\rm HI\rightarrow  H_2$
transition takes place.   We are thus able to  identify the H$_2$-rich
regions of the  interstellar medium and explore their relation to the
ongoing star formation. Our most  important findings are:  a) a 
significant dependence  of the $\rm HI \rightarrow  H_2$ transition 
and the resultant  H$_2$ gas mass on  the ambient  metallicity  and  
the H$_2$  formation  rate, b)  the important  influence of  the 
characteristic  star  formation timescale (regulating the ambient 
FUV radiation field) on the equilibrium H$_2$ gas mass and c) the
possibility  of a diffuse H$_2$ gas phase existing well beyond the 
star-forming sites where the radiation  field is low. We expect these
results to be valid in other types of galaxies for which  the 
dense and  cool HI  precursor and  the resulting  H$_2$ gas phases 
are currently inaccessible by high resolution numerical studies
(e.g.  large spirals).  Finally we implement and briefly explore 
a novel approach of using the ambient H$_2$ gas mass fraction as a 
criterion for the onset of star formation in such numerical studies.

\end{abstract}


\section{Introduction}

Stars  form in  molecular clouds,  and most  stars form  in  the large
complexes  of Giant  Molecular  Clouds (GMCs).   A  general theory  to
predict the location and amount of star formation in galaxies does not
exist yet. The  global drivers of star formation  are generally sought
in  large scale  processes  that  can form  concentrations  of HI  gas
(Elmegreen  2002),  for example  gravitational  instabilities in  disk
galaxies,  or compression  by galaxy  interactions or  mergers.  These
concentrations  of neutral gas  are then  assumed to  be the  sites of
molecular cloud  and ultimately star  formation. Numerical simulations
of  galaxies have  followed this  lead and  base their  star formation
model on  the local density  directly using a Schmidt  law \cite{MH94,
S00},  or  gravitational  instability   and  a  Jeans  mass  criterion
\cite{K92,  GI97,   BO03}.   To  date   none  of  these   efforts  has
incorporated the emergence  and evolution of the H$_2$  gas along with
its  precursor  Cold  Neutral  Medium  (CNM) HI  phase,  except  in  a
semi-empirical and somewhat adhoc fashion \cite{SC02}.  The need to do
so  in a  more  self-consistent  manner is  now  recognized from  both
numerical \cite{BO03}, and observational studies \cite{WB02}.

The  interstellar  medium  (ISM)  consists  of  gas  of  wide  ranging
properties, from cold, dense molecular  clouds to the cold and warm HI
medium (CNM,  WNM), as  well as the  hot ionized medium  intermixed in
fractal-like structures. The physical processes governing the state of
the ISM  have been progressively  identified in the last  few decades,
but  their  precise  working  is  still an  area  of  active  research
\cite{V02}.  Most  galactic-scale simulations include  only one phase,
the WNM HI \cite{K92,NW93, S00}. Some authors include more physics and
consider a two-phase  medium (Gerritsen \& Icke 1997;  Gerritsen \& de
Blok  1999).  Semi-empirical  models of  a multiphase  ISM \cite{SH02,
AB00} suffer  from the inclusion of many  poorly understood parameters
needed to describe  the interaction of the various  phases, which also
forces  such  models  to  include  serious  simplifications  (e.g.   a
constant and uniform heating of the  gas) in order to keep the problem
within the  current computing  capabilities (Semelin \&  Combes 2002).
Thus the robustness of such semi-empirical approaches in modeling real
galaxies is rather limited.   Moreover, apart from the obvious fallacy
of forming  stars out of atomic  gas, the absence of  molecular gas in
the simulations inhibits the comparison of the model galaxies with 
an  extensive  body   of  observational  data,   namely  the 
distribution of H$_2$ in galaxies  (as deduced via its tracer molecule
CO).   The latter is  observationally well-studied  \cite{EPB03, MY01,
HR01,  RTH01},  it  can  offer  additional constraints  on  the  model
galaxies, and  the same is  true for its observed  empirical relations
with other galactic components e.g.  the young stars.  For example the
Schmidt law  that links local gas  content to the  star formation rate
has been  demonstrated to  be a much  tighter relation  for molecular,
rather than HI gas \cite{WB02}.

  Some  pioneering   work  that   includes  H$_2$  in   spiral  galaxy
simulations  has been  done  by Hidaka  \&  Sofue (2002)  but they did 
not consider a  time-dependent $\rm  HI \to H_2$  transition, which  
as we will  argue in  this work,  is essential. Our time-dependent  
treatment of  the  $\rm HI\rightarrow H_2$  transition,  implemented 
here  in  an  N-body/SPH  code, can  be incorporated into other  
types of code that include  the ISM evolution {\it thus enabling  the 
resulting galaxy simulations to  track the gas phase most relevant  
to star formation, the molecular  gas.}  Our work
fills two  important ``gaps'' in the published  literature, namely: a)
it makes  a connection between galaxy-sized  simulations and molecular
cloud  theory (which  lies  behind  the emergence  of  the power  laws
observed in the  cool ISM), and b) connects  large scale instabilities
and  the appearance  of  GMC-type complexes.   The  latter allows  the
integrated study  of star formation and  the H$_2$ gas  in the dynamic
setting  of  realistic galaxy  models  that  include  effects of  e.g.
spiral   density  waves,   self-propagating  star   formation,  galaxy
interactions  and  mergers,  and  a temporally  and  spatially  varying
interstellar radiation field.  Our  presentation proceeds from a brief
exposition  of the  H$_2$ formation/destruction  theory and  the cloud
power-laws  underlying  our sub-grid  physics  assumptions,  on to  its
implementation  in  our  numerical  simulations while  discussing  the
relevant details of the code we use.  Finally we present first results
from simulations of typical  dwarf galaxies, and conclude by outlining
future work using the H$_2$-tracking numerical models.

\section{H$_2$ formation and destruction}

The  formation  of H$_2$  in  galaxies  has  already been  studied  by
Elmegreen  (1989,   1993),  where   the  important  role   of  ambient
metallicity and  pressure has  been described.  Subsequent  efforts to
incorporate  his approach  into  analytical models  of galactic  disks
\cite{HSA95}  or   numerical  simulations  \cite{HS02}   have  adopted
stationary  models,  namely  once  an  H$_2$  formation  criterion  is
satisfied  the  $\rm  HI\rightarrow  H_2$  transition  is  set  to  be
instantaneous.  The time-dependence  of several factors affecting this
transition (e.g.  the  ambient H$_2$-dissociating FUV radiation field)
and the various gas heating and cooling processes were not considered,
seriously  restricting  the  ability  of  such  models  to  track  the
evolution  of  the ISM  and  particularly its  H$_2$  gas  phase in  a
realistic  manner.   This  is  because the  conditions  affecting  the
HI$\leftrightarrow$H$_2$   equilibrium  can   vary   over  timescales
comparable or shorter than that  needed for the latter equilibrium to
be reached.

 Dust  grains, when  present, are  the  sites where  H$_2$ forms.  Its
formation  rate $\rm  R_f$  for  gas with  temperature  $\rm T_k$ 
 and metallicity $\rm Z$ can then be expressed as

\begin{equation}
\rm R_f = \frac{1}{2} \sigma_d \langle v_H \rangle
 S_H \gamma_{H_2}=3.5\times 10^{ -17} \mu \, Z\,  \left(\frac{T_k}{100\ K}\right)^{1/2} \,
 S_H \gamma_{H_2}\, cm^3\  s^{-1},
\end{equation}
 
\noindent
(e.g.  Hollenbach,  Werner \& Salpeter  1971; Cazaux \&  Tielens 2002,
2004).  This equation  expresses the H$_2$ formation rate  as the rate
at  which HI  atoms with  mean  velocities $\rm  \langle v_H  \rangle$
collide with  interstellar dust particles with  effective surface $\rm
\sigma_d$, multiplied  by the probability $\rm  S_{H}(T_k)$ they stick
on  the grain  surface  and the  probability  $\rm \gamma_{H_2}$  they
eventually  form an  H$_2$  molecule that  detaches  itself from  the
grain.   The grain  surface $\rm  \sigma_d$  (and thus  $\rm R_f$)  is
assumed   to  scale   linearly   with  metallicity   Z,  namely   $\rm
\sigma_d=\sigma_g  n_g/n =  4.9 \times  10^{-22} Z\  cm^2$  (with $\rm
\sigma_g$ the grain geometric cross section and with the ratio of grain 
density to total hydrogen density $\rm  n_g/n \propto  Z$). Laboratory
experiments usually constrain only the $\rm S_H \gamma _{H_2}$ product
for various  temperature domains  rather than provide  information for
each function  separately (e.g.  Pirronello  et al. 1997; Katz  et al.
1999).   The  theoretical study  of  Buch  \&  Zhang (1991)  yields  a
function   $\rm   S_{H}=   [1+(k_B   T_k/E_{\circ})]^{-2}$   (with   a
characteristic  energy scale  $\rm E_{\circ}/k_B\sim~100$~K obtained
from fitting  results from  molecular dynamics simulations)  valid for
$\rm T_k\lesssim 300$  K, which we adopt in the  present work (but see
also Cazaux \& Spaans 2004 for the most recent views).

We  incorporate all  the  uncertainties  of $\rm  R_f$  into a  single
parameter  $\mu$.  These  uncertainties stem  mainly from  the adopted
probability functions  and the effective surface  for H$_2$ formation.
For example, the effective H$_2$  formation surface may be bigger than
implied  by the visual extinction  cross-section $\rm \sigma_d$.
The canonical formation rate $\rm R_f=3 \times 10^{-17}\ cm^3\ s^{-1}$
(Jura  1974, 1975)  for typical  CNM HI  gas ($\rm  T_k\sim  100\ K$),
corresponds  to $\mu  =3.5$ but  values of  $\mu \sim  17-18$  are not
excluded.  Early  suggestions for such high $\mu$  values emerged from
the  study of  rovibrational  infrared lines  of  H$_2$ in  reflection
nebulae (Sternberg 1988), and observations of intense H$_2$ rotational
lines with  the Infrared Space Observatory  (ISO) in photodissociation
regions \cite{HBV00, LEJ02}.

The timescale associated with  H$_2$ formation is then (e.g.  Goldshmidt \&
Sternberg 1995)

\begin{equation}
\rm  \tau_{f} = (2 n_1 R_f)^{-1}=5\times 10^7
\left(\frac{T_k}{100\ K}\right)^{-1/2}\left(\frac{n_1}{10\
cm^{-3}}\right)^{-1} \left[\mu \, Z\, S_{H}(T_k)\right]^{-1}\ yrs,
\end{equation}

\noindent
where $\rm  n_1 $ and  $\rm T_k$ are  the HI density  and temperature.
For $\rm n_1 \sim 50$ cm$  ^{-3}$ and $\rm T_k \sim 100$~K, typical of
the Cold  Neutral Medium HI gas  out of which H$_2$  clouds form, $\rm
\tau _{f}\sim  10^7$ yrs.  This is  comparable to the  timescales of a
wide  variety of  processes  expected to  fully  disrupt or  otherwise
drastically   alter  typical  molecular   clouds  and   their  ambient
environments.   Some of the  most important  ones are  star formation,
with  the disruptive  effects  of O,  B,  star clusters  \cite{BGP77},
turbulent dissipation \cite{SOG98,  MC98}, and inter-cloud clump-clump
collisions  \cite{BS80}.   The {\it  mean}  FUV  field  driven by  the
evolution  of  continously forming  stellar  populations throughout  a
galaxy evolves over similar  timescales \citep[e.g.][]{PHM03},
and the same seems to be the case for the ambient pressure environment
and its  perturbations by passing SN-induced shocks  \citep[e.g.][]{WMHT03}.  
{\it Hence a  realistic model of the HI$\rightarrow $H$_2$
transition in galaxies must be time-dependent.}

With this necessity in mind, we will first discuss the equilibrium 
fraction of molecular gas in the ISM, before presenting a 
time dependent version of the model.  

\subsection{The equilibrium molecular fraction}

The equilibrium molecular gas  fraction  \emph{per}  cloud,  under a  given
ambient   FUV   field,   can   be   estimated   by   considering   the
formation/destruction   equilibrium  for   a   plane  parallel   cloud
illuminated  by  a radiation  field  $\rm  G_{\circ}$. In  equilibrium
formation  balances destruction, namely

\begin{equation}
\rm R_f n n_1 = G_{\circ} k_{\circ} f(N_2) e^{-\tau} n_2,
\end{equation}

\noindent
which must hold  at any depth.  The densities $\rm  n_1$ and $\rm n_2$
denote the HI and H$_2$ densities, while the total hydrogen density is
$\rm  n=n_1+2 n_2$.   The  H$_2$ dissociation  rate  $\rm k_{\circ}  =
4\times 10^{-11}\  s^{-1}$ is normalized  for an ambient FUV  field in
units of  the Habing \cite{H68} field value  ($\rm G_{\circ}=1$).  The
factor $\rm  f(N_2)$ is  the normalized H$_2$  self-shielding function
which  describes the  decrease in  dissociation  rate due  to the  FUV
absorption  by  the molecular  column  $\rm  N_2$.  Furthermore,  $\rm
\tau=\sigma  (N_1+2  N_2)$ is  the  FUV  optical  depth due  to  grain
extinction,   with   a   dust   FUV  absorption   cross-section   $\rm
\sigma=\xi_{FUV} \sigma_d$  ($\rm \xi_{FUV}=2-4$).  Eq.  3 can be
converted to a separable differential  equation for $\rm N_1$ and $\rm
N_2$  \citep[e.g.][]{GS95}  which, when  integrated  by  parts  for  a 
uniform cloud, yields a total HI column density of

\begin{equation}
\rm N_{tr}(HI) =\frac{1}{\sigma} ln\left(1+\frac{G_{\circ}
 k_{\circ}}{ n R_f} \Phi\right).
\end{equation} 

\noindent
This is  the total steady-state  HI column density resulting  from the
FUV-induced H$_2$ photodissociation  of a one-side illuminated uniform
cloud, and can be taken as  a measure of the total `thickness' of
  the  HI   layer.  The   factor  $\Phi$  is   an  integral   of  the
self-shielding function  over the  H$_2$ column, which  encompasses the
details   of    H$_2$   self-shielding.    For    $\rm   f(N_2)   \sim
(N_2/N_{ch})^{-k}$ it is

\begin{equation}
\rm \Phi=2^{k-1}\left(N_{ch}\sigma\right)^k \int _{0} ^{\infty} x^{-k} e^{-x}dx
=6.6\times 10^{-6} \sqrt{\pi}\, Z^{1/2}\, \xi ^{1/2} _{FUV}.
\end{equation}

\noindent
Its numerical value was obtained for $\rm k=1/2$ and a characteristic 
column density of $\rm N_{ch}=1.75\times 10^{11}\ cm^{-2}$ \cite{J74}.

 The equation of the  $\rm HI\leftrightarrow H_2$ balance still yields
an analytical expression for  $\rm N_{tr}(HI)$ for clouds with density
profile: $\rm n(r)=n_e (r/R)^{-1}$ ($\rm R$: the cloud radius). Such a
profile results from cloud models using a logotropic equation of state
\cite{MP96}, hence we  will refer to them as  "logotropic clouds," but
the reason  for considering  such profiles here  is to obtain  a rough
representation of sub-resolution,  non-uniform, GMC density structure.
The corresponding expressions for the equilibrium transition column 
density etc. for logotropic clouds are given in the Appendix.

The visual extinction corresponding  to a transition column $\rm N_{tr}(HI)$ 
is given by (using also the expressions for $\rm R_f$, $\Phi $ from Eqs 1 and 5)

\begin{equation}
\rm A_v ^{(tr)}=1.086 \xi ^{-1} _{FUV}
\ln\left[1+\frac{G_{\circ}}{\mu
 S_{H}(T_k)}\left(\frac{\xi _{FUV}}{Z T_k}\right)^{1/2} 
\left(\frac{n_e}{135\ cm^{-3}}\right)^{-1}\right].
\end{equation}

\noindent
For a spherical constant density cloud the H$_2$ gas mass fraction contained 
``below''  the HI  transition zone  marked by $\rm  A_v ^{(tr)}$ is 
then given by

\begin{equation}
\rm f_m \equiv \frac{M(H_2)}{M_c}=\left(1 - \frac{4 A_v ^{(tr)}}{3 \langle A_v
 \rangle}\right)^{3},
\end{equation}

\noindent
where  $\rm \langle  A_v \rangle$  is the area-averaged extinction  
of the cloud measured by an outside observer ($\rm f_m=0$ for  
$\rm  A_v ^{(tr)} \ge 3/4 \langle  A_v \rangle$) .

 We consider  $\rm f_m$  to be  a measure of  the local  molecular gas
content of  the ISM,  the implicit assumption  being that most  of the
mass of the  cool HI/H$_2$ gas can be ultimately  traced to such cloud
structures.   For CNM  HI and  the resulting  H$_2$ clouds  this  is a
reasonable assumption and in certain  models it is indeed predicted to
be the final configuration of any  initial warm ($\sim 10^4$ K) HI gas
mass which quickly  cools and fragments in an  environment of constant
``background'' pressure \cite{C87, CP87}.

\subsection{Sub-resolution physics: the cloud size-density relation}

From  the previous  discussion it  becomes apparent  that in  order to
calculate $\rm f_m$ for a region of the ISM we need  an expression
  for  the  typical local  mean  cloud  extinction  $\rm \langle  A_v
\rangle$.  The crucial  assumption we make here to  derive an estimate
for $\rm \langle A_v \rangle$  is that all unresolved cloud structures
where  the HI/H$_2$ transition  takes place  obey the  widely observed
density-size  power law  \cite{L81}.  The  bulk  of the  H$_2$ gas  is
indeed observed to reside in Giant Molecular Clouds, the sites of most
star  formation activity  in galaxies,  and these  are observed  to be
virialized  entities  \cite{L81,   E89,  ROS03}.   Their  density/size
scaling  relation can  be derived  from an  application of  the virial
theorem to clouds under a given boundary pressure $\rm P_e$ \cite{E89}
and a universal (velocity  dispersion)-(cloud size) relation.  Then it
can be shown that the average H density is given by

\begin{equation}
\label{eq_nmean}
\rm  \langle n \rangle  = n_{\circ} \left(\frac{P_e/k_B}{10^4\ K\ cm^{-3}}\right)^{1/2}
 R^{-1} _{pc},
\end{equation}

\noindent
where $\rm R_{pc}$  is the cloud radius in parsecs.   The range of the
normalizing constant  $\rm n_{\circ}$  (cm$ ^{-3}$) can  be determined
theoretically by choosing a range of plausible polytropic solutions to
the cloud density profiles \cite{E89},  or from the {\it observed} n-R
relations found  for galactic molecular clouds.   The latter approach,
after using the scaling relation $\rm n(H_2) = 1700 R^{-\alpha} _{pc}\
cm^{-3}$ reported by Larson (1981)  ($\rm \alpha \sim 1$), yields $\rm
n_{\circ}\sim  1520$~cm$ ^{-3}$.  In  estimating $\rm  n_{\circ}$ from
the observed  n-R scaling relation we  took into account  that for the
galactic   midplane   molecular  clouds   the   outer  pressure   $\rm
P_{\circ}/k_B\sim 10^4\ K\ cm^{-3}$, but the boundary pressure on the
molecular  part  of  the  cloud  $\rm P_e=P_{e,m}\sim  5  P_{\circ }$
\cite{E89}.

It  is worth  mentioning  that the  invariance  of the  linewidth-size
relation  $\rm \Delta  V(L)  \sim V_{\circ}  (L/pc)^{1/2}$ (which  for
virialized clouds yields  Eq. \ref{eq_nmean}) has been recently  
verified over an extraordinary range of conditions as well as for 
the environments {\it within} clouds  \cite{HB04}.  This  lends 
additional support to the  use of the  n-R scaling law  as a 
non-evolving element  of our assumed sub-grid  physics. These 
relations are expected  to break down only at very small scales where 
linewidths become dominated by thermal rather than macroscopic motions. 
For typical CNM temperatures setting $\rm \Delta  V(L)\sim \Delta V_{th}$
yields $\rm L\la  0.5$~pc, much smaller  than   the  spatial  resolution 
achievable  in  galaxy-size simulations like ours.

The average extinction of a cloud with radius R as measured
 from  outside is,

\begin{equation}
\rm \langle A_v \rangle = \frac{2 k_g \langle n \rangle Z R}{N_{A_v=1}(H)},
\end{equation}

\noindent
where  the geometric factor  $\rm k_g  = 2/3$  for a  spherical cloud.
After substituting  $\rm \langle n  \rangle$ from Eq. \ref{eq_nmean}
and setting $\rm N_{A_v=1}(H)=1.85\times 10^{21} cm^{-2} $ we obtain

\begin{equation}
\rm \langle A_v \rangle = 0.22 Z \left(\frac{n_{\circ}}{100\ cm^{-3}}\right)
\left(\frac{P_e/k_B}{10^4\ cm^{-3} K}\right)^{1/2}.
\end{equation}

The cloud boundary  pressure is due to thermal  as well as macroscopic
motions  (Elmegreen 1989), and assuming  the latter  to be  isotropic, 

\begin{equation}
\label{eq_pext}
\rm P_e = n_ek_B T_k + \frac{1}{3} \rho _e \sigma_{vel}^2 = \left[1.085 T_k
+54 \left(\frac{\sigma_{vel}}{km\ s^{-1}}\right)^2 K \right] n_e k_B,
\end{equation}

\noindent
where $\rm  \sigma_{vel}$ is the 3-dimensional  velocity dispersion of
the gas due solely to macroscopic motions.  From Eqs 7 and 10 it is
obvious that  high pressure  gas will tend  to be molecular,  a result
already  known from  the  analytical/steady-state models  of the  $\rm
HI\rightarrow H_2$ transition in galactic disks \cite{E93, HSA95}.

We  can now  compute  the   equilibrium  $\rm  f_m$ values  for
interstellar gas  for a given $\rm (n,  T_k)$ sampling the local conditions
in the ISM by using Eq. 7 (or Eq. B3 for logotropic clouds). Input 
parameters are the radiation  field $\rm G_0$, 
metallicity Z, velocity dispersion $\rm  \sigma_{vel}$ and our H$_2$ 
formation rate parameter $\mu$.  In  Figure 1 we  show the resulting 
equilibrium  fraction $\rm f_m$ for  a number  of typical values  of 
these parameters.   The true equilibrium molecular gas content will 
most likely be higher than $\rm f_m$ since the real FUV field  
incident on GMCs is not radial and thus falls off more  rapidly inside 
the absorbing cloud  layers. Higher gas densities  expected to  exist
deeper in  the  clouds as  part of  any spatial and density 
substructure/clumping in the ISM will act to raise the  true $\rm  f_m$. 
In  that respect  logotropic  clouds are  more realistic than those 
with uniform density.

\subsection{Time-dependence of the HI/H$_2$ equilibrium}

The time-dependent  HI/H$_2$ transition  can be approximated  from the
solution  of Equation  22  of Goldshmidt  \&  Sternberg (1995),  which
describes the evolution of the total HI column $\rm N_{tr}(HI)$ in the
$\rm HI  \to H_2$ transition layer  of a plane parallel  slab of H$_2$
gas with an impinging dissociating radiation field:

\begin{equation}
\rm \tau_f \frac{d \sigma N_{tr}(HI,t)}{d t}=r_{dis} e^{-\sigma N_{tr}(HI,t)} - 
 \sigma N_{tr}(HI, t),
\end{equation}

\noindent
where $\rm \tau_f=1/ (2 n R_f) $ (the H$_2$ formation timescale in the
fully atomic part of the  cloud), and the  dimensionless quantity
$\rm r_{dis}$ quantifies the  local balance between H$_2$ dissociation
and formation and is given by

\begin{equation}
\rm r_{dis} \equiv \frac{G_{\circ} k_{\circ}}{n R_f}\Phi = 2.67 \frac{G_{\circ}}{\mu S_H(T_k)}
 \left(\frac{\xi _{FUV}}{Z\, T_k}\right)^{1/2} \left(\frac{n}{50\ cm^{-3}}\right)^{-1}.
\end{equation}

\noindent
The  assumption underlying  the  validity of  Eq. 12 is that  of a  sharp
 HI/H$_2$ transition layer separating the slab into a fully atomic and
 a fully molecular section.  In the case of steady state Eq. 12 reduces to

\begin{equation}
\rm r_{dis} e^{-\sigma N_{tr}(HI)}= \sigma N_{tr}(HI),
\end{equation}

\noindent
which does not yield $\rm N_{tr}(HI)$  as expressed in Eq. 4 since the  
latter does not involve the  approximation of a sharp HI/H$_2$ 
transition made here.

In order  to see at what  step of deducing  Eq. 4 the assumption  of a
sharp HI/H$_2$  transition layer would yield  the result in  Eq. 14 we
can use Eq. 2 from Goldshmidt and Sternberg (1995)

\begin{equation}
\rm R_f n dN(HI)=G_{\circ} k_{\circ} f[N(H_2)]
 e^{-2\sigma N(H_2)} e^{-\sigma N(HI)} dN(H_2).
\end{equation}

\noindent
For a  sharp HI/H$_2$ transition  zone not much  HI can exist  in that
zone to significantly  add to the total HI  transition column density.
Hence, when  integrating the last  equation by parts,  $\rm e^{-\sigma
  N(HI)}$ can be replaced  by $\rm e^{-\sigma N_{tr}(HI)}$. The latter
is then  treated as  a constant of  integration over the  H$_2$ column
density and  kept out of the integral, yielding Eq.  14.  We find
the steady-state  solutions provided by Eq.  14 to differ  $\la 30\% $
with  those  provided  by  the   more  accurate  Eq.   4  while  other
uncertainties  of  our   model  (e.g.  the  effects  of   CNM  HI  gas
substructure) are expected to be more important.  The sharpness of the
HI/H$_2$ transition zone  ($\rm \Delta A_v\la 0.1$ for  Z=1) is due to
the extreme  self-shielding properties of  the H$_2$ molecule  and has
been  known since  the early  theoretical models  of Photodissociation
Regions (PDRs)  (Hollenbach, Takahashi \& Tielens  1991; Hollenbach \&
Tielens 1999).

A  fully  time-dependent  treatment  of  the  HI/H$_2$  transition  is
provided by solving Eq.  12 and then using $\rm A^{(tr)}_{v}=\xi ^{-1}
_{FUV} \left[ \sigma N_{tr}(HI,t)  \right] $ to find the corresponding
$\rm  f_m(t)$ from  Eq.  7.  Although the  approximation  of a  narrow
transition layer is not strictly valid, the time dependent solution of
the full integro-differential equation would  be too costly for use in
numerical  simulations, and  also  the uncertainties  inherent in  our
model  would not  justify the  additional expense  for the  more exact
solution. In Figure \ref{GSfig} we compare the solution of Eq. 12 with
the full solution as calculated by Goldshmidt \& Sternberg (1995). The
main  difference is  that  our  solution for  the  $\rm N_{tr}$  grows
roughly $30\%$ slower, which gives an  effect   similar to an error of
$30\%$ in e.g. the grain opacity or H$_2$ formation rate parameter.

\subsection{Collisional destruction of H$_2$ at high temperatures}

The preceding  description of the  HI/H$_2$ balance is not  valid 
  for gas much warmer than its CNM phase where collisional rather than
  FUV-induced  destruction of H$_2$  dominates. Under  such conditions
  $\rm  R_f\sim  0$  since once  $\rm  T_k  \ga  1000$ K,  the  H$_2$
formation effectively becomes zero \cite{CT04} and any remaining (from
the FUV destruction) H$_2$ will eventually be collisionally destroyed,
especially at  temperatures $\rm T_k\ga 3 \times  10^3$.  Moreover, at
such  high  temperatures  the   cloud  scaling  laws  (and  hence  the
expressions  used to  estimate $\rm  f_m$)  no longer  hold since  the
observed ``cloud'' linewidths are  then almost purely thermal and thus
no longer scale with size.

In  principle the collisional  destruction of  a purely  molecular gas
will first  be dominated by H$_2$-H$_2$ collision  processes and later
on by the more efficient H$_2$-HI collisions. If we look at the collisional
destruction coefficients for  these two processes we see  that for the
H$_2$ collisions to be more important the condition

\begin{equation}
\rm  \frac{n_2}{n_1} > 10,
\end{equation}

\noindent
must be  satisfied at $\rm  T_k\sim 3\times10^4$ K (Martin,  Keogh, \&
Mandy 1998).  In the WNM gas phase such a high percentage of remaining
H$_2$  (remaining   from  the  dominant   FUV  destruction  mechanisms
operating in the CNM and the CNM-to-WNM phases) is unlikely.  We
thus consider  only the HI  collisional destruction term  as important
and then the H$_2$ density will be controlled by

\begin{equation}
\rm  \frac{dn_2}{dt} =-\gamma_1(T_k) \left(n-2n_2\right) n_2,
\end{equation}

where $\gamma_1$ is the H$_2$-HI collisional destruction coefficient. 
The (analytic) solution of this will be employed for the high 
temperature domain. Here we must mention that we assumed no FUV-induced 
HI/H$_2$ spatial segregation in the  WNM phase.  If there  is some 
remnant spatial  segregation H$_2$ would be destroyed  by the much 
less effective  H$_2$ collisions, thus our  assumption most  likely 
overestimates the  level of  collisional H$_2$ destruction in the 
WNM phase.

\section{Implementation}
 
We implement  a module incorporating the  aforementioned ``recipe'' to
track  the  $\rm HI\leftrightarrow  H_2$  gas  phase  interplay in  an
N-body/SPH code  for galaxy simulations  that includes a model  of the
neutral gas  phases. N-body/SPH  codes are in  routine use  within the
astrophysical  community  as  tools  to  explore  problems  in  galaxy
evolution  and formation,  and  their full  description  can be  found
elsewhere  \cite{HK89,   M92}.  We   use  the  new   conservative  SPH
formulation of Springel \& Hernquist (2002).  The artificial viscosity
used is the standard Monaghan (1992) viscosity.

The code we use is a major  upgrade of Gerritsen (1997) and Gerritsen 
\& Icke  (1997), mainly in its description of  the evolution  of the 
stellar and gaseous components. A complete description  can be 
found in Pelupessy (2005), here we will highlight the aspects most 
relevant.

The  main   feature  making   this  code  especially   attractive  for
incorporating  the  $\rm  HI/H_2$  phase transition  is  its  detailed
modeling of the  ISM, including physics of the WNM  and CNM HI phases,
as well as the modelling  of star formation and feedback.  This allows
us  to conduct  high resolution  simulations following the ISM gas  as it
cools down  to temperatures of  $\rm T <  100$ K and densities  $\rm n
\sim 100$  cm$^{-3}$, conditions  expected to be  typical of  the $\rm
HI\rightarrow H_2$ transition phase.

Note that the addition of our HI/H$_2$ model does not add a major 
computational burden to the code: in practice we find that $< 1 \%$ of 
CPU time is taken up by the molecule formation routines. Furthermore 
the necessary calculations (solving Eq. 12 using an implicit integration 
scheme) scale linearly with the number $N$ of particles, whereas the 
time consuming parts of the simulation (gravity and neighbour 
searching, which take up $\approx 60\%$ of CPU time for the simulations 
presented here) scale as $\rm N~log~N$ for an N-body/SPH code\cite{HK89}.
 
\subsection{The neutral ISM model}

Our model for  the ISM is similar to the equilibrium model of Wolfire
et al. (1995,  2003) for the neutral phases and  was extended from the
more  simplified models  from  Gerritsen \&  Icke  (1997) and  Bottema
(2003).   We  consider  a   gas  with  arbitrary  but  fixed  chemical
abundances  $\rm X_i$,  scaled to  the target  metallicity  from solar
abundances. We solve  for the ionization and thermal  evolution of the
gas. The  thermal evolution is solved with  a predictor/corrector step
as in \citep{HK89}.  The various processes included in  the ISM model
are given in Table~1. The main differences with the work of Gerritsen
\&  Icke (1997)  and Bottema  (2003) are  the following:  we  use more
accurate cooling,  that is calculated in accordance  with the chemical
composition,  we  solve  for  the  ionization  balance  (albeit  still
assuming  ionization equilibrium)  and we  use the  full photoelectric
heating efficiency as given in Wolfire et al. (1995).
Indicatively,  in   Figure  \ref{ismfigm}  we   plot  the  equilibrium
temperature, ionization fraction, heating (=cooling) and pressure as a
function of density.  There it can  be seen that as density varies the
equilibrium  state of  the gas  changes from  a  high temperature/high
ionization  state  ($\rm T=10^4~K$,  $\rm  x_e  \approx  0.1$) at  low
densities, to  a low temperature/low ionization  state ($\rm T<100~K$,
$\rm x_e<10^{-3}$) at  high densities.  In between there  is a density
domain where the negative slope of the P-n relation indicates that the
gas is  unstable to isobaric pressure variations,  the classic thermal
instability \cite{F65}.  The shape of these curves and hence the exact
densities  of  the thermal  instability  vary  locally throughout  the
simulation,  influenced by  the  time-varying UV  radiation field  and
supernova heating.  We set a  constant cosmic ray ionization rate $\rm
\zeta_{CR}$ throughout  the galaxy, assumed to  be $\rm \zeta_{CR}=3.6
\times 10^{-17}~s^{-1}$.

Although we  will consider models  of different metallicities  we will
not consider the effects of abundance gradients or enrichment here. In
general the  abundance gradients observed in dwarf  galaxies are small
\cite{PE81}, so this  is a reasonable approximation. The  fact that we
keep  the  metallicity  constant  in  time means  that  the  model  as
presented  here is  not  yet  suitable to  follow  the evolution  over
cosmological  timescales or  to simulate  very low  metallicity dwarfs
(for the simulations  presented here the evolution in  Z would be less
than $10\%$ over the course of the simulation, even if no metals 
  were to be lost to the intergalactic medium).
          
\subsection{Star formation and feedback}

The coldest and densest phase in our model is best identified with the
CNM,  where GMCs most  likely form  and remain  embedded.  We will 
use the simple  prescription for  star formation  of Gerritsen  
\&  Icke (1997) as our standard star formation model. It was shown to reproduce  
the star formation  properties of ordinary spiral galaxies and it is based on 
the assumption  that the star formation process is governed by gravitational
instability.  A gaseous region is considered unstable to star formation if 
the local Jeans mass $\rm M_J$,

\begin{equation}
\rm M_J \equiv \frac{\pi \rho}{6} \left( \frac{\pi s^2}{ G \rho} \right)^{3/2} <
M_{ref}
\end{equation}
 (with  s the  sound speed), the assumption being that structure on 
 a mass scale $\rm M_{ref}$ is present in the ISM. 
 Once a region is dense and cold enough that (18) is fullfilled, the rate 
 of star formation rate is set to scale with the local free fall time 
 $\rm t_{ff}$:

\begin{equation}
\rm \tau_{sf}=f_{sf} t_{ff}= \frac{f_{sf}}{\sqrt{4 \pi G \rho}}
\end{equation} 

\noindent
The  delay  factor $\rm  f_{sf}$  accounts  for  the fact  that  cloud
collapse  is inhibited by  either small  scale turbulence  or magnetic
fields  \cite{MK04, SAL87}.  Its  value is  uncertain and  we consider
values  $\rm f_{sf}  = 2.5  - 20$.   The actual  implementation  of star
formation  works  as follows:  once  a gas  particle  is  found to  be
unstable according  to the  Jeans mass criterion  it can spawn  a star
particle with a mass one eighth of  the mass of a gas particle (thus a
given  gas particle  can produce  at most  eight star particles).  The
process is governed by either drawing  from a Poisson distribution 
such that  the  local rate  of  star  formation agrees  with  Eq.  19 
in  a stochastic sense, or by imposing a fixed delay time.

 Our basic star formation  recipe does  not depend  explicitly on  the local
H$_2$ fraction, so while  the gravitational  collapse that  induces star 
formation  may also  precipitate  H$_2$  formation, it is  nevertheless 
independent of whether the  gas is atomic  or molecular. However,  our model now 
allows for a direct link between star formation and the presence of 
molecular gas. We can do this by setting a local threshold for $\rm f_m$ 
above which star formation can proceed. As we will discuss in section 4.2.2 
this may be more realistic since in such a scenario the values of the 
delay factor are no longer assumed a priori but are instead implemented
in a more physical fashion by the ``delay'' associated with the final 
chemical/ thermodynamic evolution of a gas cloud before star formation.

In order  to determine  the local FUV  field used for  calculating the
photoelectric heating  and the  H$_2$ destruction rates,  we determine
the time-dependent FUV  luminosity of the stellar particles.  We do so
by  following  their age and, since  star particles  represent
  stellar associations rather than  single stars, by using Bruzual \&
Charlot (1993  and updated) population  synthesis models. We  assume a
Salpeter  initial  mass  function  (IMF)  with  cutoffs  at  $0.1  \rm
M_{\odot} $ and $\rm 100\rm M_{\odot }$. In the present work we do not
account  for dust extinction  of UV  light, except  that of  the young
stars shrouded  in their natal cloud:  for a young  stellar cluster we
decrease  the amount  of  UV extinction  from  75\% to  0\%  in 4  Myr
\cite{PHM03}.

Feedback from stellar winds and supernovae is essential for regulating
the  physical conditions of the ISM.  While the mechanical energy
output  of  stars  is reasonably  well  known,  it  has proven  to  be
difficult   to  include  it   completely  self-consistently   in  
  galaxy-sized simulations of  the ISM.  The reason for  this is that
the  effective energy of  feedback depends  sensitively on  the energy
radiated away  in thin shells  around the bubbles created.  This will
mean that the effect of  feedback cannot be tracked in a straightforward 
manner unless prohibitively high resolution is  used. In SPH codes there 
have been conventionally two ways to account for this: 
by changing the thermal energy input
and by acting on particle velocities.  Both are unsatisfactory, as the
thermal  method suffers  from overcooling  \cite{K92} and  the kinetic
method  seems  to   be  too  efficient  in     stirring  the  ISM
\cite{NW93}.  Here  we use a new  method based on the  creation at the
site of young stellar clusters  of a {\it pressure particle} that acts
as a normal SPH particle in the limit of the mass of the particle $\rm
m \rightarrow 0$, for constant  energy.  For the energy injection rate
(which  assumes that  the energy  injection takes  place continuously,
without distinguishing between stellar winds and supernovae) we take
\begin{equation}
\rm  \dot{E}=\epsilon_{sn}  n_{sn} E_{sn}/   \Delta    t,
\end{equation}  
with $\rm  E_{sn}=10^{51}$ erg the energy liberated  per supernova, an
efficiency parameter $\rm  \epsilon_{sn}=0.1$, the number of supernova
per mass  of new stars  formed $\rm n_{sn}=0.009$ per  $\rm M_{\odot}$
(appropriate for the IMF adopted here) and $\rm \Delta t=3 \times 10^7$
yr.  The efficiency  $\rm \epsilon_{sn}$ thus assumes that 90\% of
the initial supernova energy is radiated away in structures not 
resolved by our simulation. This value has been found in detailed
simulations of the effects of supernovae and stellar winds  on the 
ISM \cite{SFPT96, Tetal98}, and is also used in  other simulations 
of galaxy evolution  \cite{SC02, SH02, BCCL00}. Moreover in a study
by Pelupessy et al. (2004) the adopted feedback strength has been shown 
to be consistent with the observed scatter in the star formation 
properties of isolated dwarf galaxies. Details of the feedback model, including
implementation features and observational constraints can be found in
Pelupessy et  al. (2004) and Pelupessy (2005). The most 
important effect of SN  feedback for our model of H$_2$ formation is, 
apart  from the regulating  effect of feedback on  star formation, the 
introduction of extra  dynamical pressure  (Eq.  11)  through the 
increases in local velocity dispersion.

\subsection{H$_2$ formation}

For our H$_2$ formation model we follow the same philosophy as for the
star formation model: unresolved structure is assumed to be present at
the SPH  particle positions.  In  this case the  underlying structures
are assumed to  conform to Eq.  10 expressing  the mean extinction and
derived  from the  observed  density/size scaling  relation (Eq.   8).
However  the  latter  is  not  expected to  be  valid  throughout  our
simulation  domain.  Regions where  the density/size  relation becomes
inapplicable are those with low density and pressure where it predicts
very large  cloud sizes and  the resulting photo-destruction  of H$_2$
proceeds  very slowly.   Such regions  contain  WNM HI  gas where  the
pressure is mostly  thermal and then Eq. 8 (along with  $\rm P_e = n_e
k_B  T_k$)  yields $\rm  R_{pc}\propto  T_k  P_e  ^{-1/2}$, which  for
typical  WNM  conditions   corresponds  to  kpc-size  ``clouds''.   To
circumvent this problem we modify Eq. 8 for $\rm P_e< P_{trans}$ as:

\begin{equation}
\rm  \langle n \rangle  = n_{\circ} 
\left(\frac{P_{trans}/k_B}{10^4\ K\ cm^{-3}}\right)^{-1/2}
 \left(\frac{P_e/k_B}{10^4\ K\ cm^{-3}}\right)
 R^{-1} _{pc},
\end{equation}

\noindent
where we take $\rm P_{trans}  \approx 1000$ K cm$^{-3}$.  This is just
a convenient patch, which however in low pressure environments  does
scale  the ''cloud''  mean density  as  $\rm \propto  P_e$, found  for
diffuse  clouds \cite{E93}.   This minor  modification now  allows the
application of our subgrid model  in all ISM conditions present in the
simulations.

For the  macroscopic pressure $\rm P_e$  in Eq.  11 we  need the local
velocity dispersion  $\rm \sigma_{vel}$.  For this we  take the formal
SPH estimate

\begin{equation}
 \rm \sigma_j^2 = \sum_i \frac{m_i}{\rho_j} (v_i-\langle v \rangle_j)^2 W( |r_{i j}|, h_j) 
\end{equation}

\noindent
with $\rm v_i$ and $\rm  m_i$ the particle velocities and masses, $\rm
\langle  v \rangle_j$  the local  bulk velocity.   A number  of subtle
issues are connected  with the choice of the  dispersion.  Equation 22
describes the inter-particle velocity dispersion, while the dispersion
that enters Eq. 11 is  really an intercloud velocity dispersion.  Even
if we assume  these to be equivalent, the problem  remains that Eq. 22
expresses a velocity  dispersion on scales of the  local SPH smoothing
length  $\rm h_i$.  We can  try to  account for  this by  scaling $\rm
\sigma_j$, using for example the relation for Kolmogorov turbulence,

\begin{equation}
\rm \sigma'_j = \sigma_j \left( \frac{h_j}{R_{cloud}} \right)^{-1/3},
\end{equation}

\noindent
but we found this to have rather little influence on the resulting 
H$_2$ formation.

Depending on  the cloud model,  we use Eqs  12 or B5 when  $\rm T_k\la
1000\ K$ to  track the evolution of $\rm  A_{v}^{(tr)}$, and thus $\rm
f_m$, during  a simulation  timestep dt.  For  $\rm T_k>1000\ K  $ and
$\rm R_f=0$ Eq. A3 describes the photodestruction of clouds, while for
$\rm  T_k>3000\ K$,  Eq.  17  is used  to approximate  the collisional
destruction process  of the  remnant molecular~gas.  The  density that
enters those equations is assumed  to be the mean density $\rm \langle
n \rangle$ given by the SPH  density at the particle position, and the
temperature  the particle  temperature  (both taken  constant for  the
timestep).  The radiation field is calculated from the distribution of
stars,  where the  assumption is  that extinction  from dust,  or from
molecular  clouds is  not important  (apart from  extinction  from the
natal  cloud).   After we  have  evolved  $\rm  A_{v}^{(tr)}$ for  the
timestep $\rm  dt$ the resulting $\rm  f_m$ (Eqs 7,  B3) is calculated
and assigned to the SPH particle for the duration of the next timestep
(where  it can  be  used to  calculate  e.g.  cooling).   At the  next
timestep  {\it the  last  $\rm f_m$  value  is retained}  and used  to
calculate the initial $\rm A^{(tr)} _{v}$, given the new average cloud
extinction $\rm \langle A_v \rangle$,  which may have changed from the
previous time-step, if for example the pressure $\rm P_e$ has changed.

The choice to  keep $\rm f_m$ rather than  $\rm A_{v}^{(tr)}$ constant
during any variations of $\rm \langle A_v \rangle $ is dictated by the
decision to  assign all variations  of $\rm f_m$ to  the FUV-regulated
HI/H$_2$ gas phase interplay, as tracked  by Eqs 12 and B5 rather than
to any  instantaneous effects.  Indeed, if  $\rm A_{v}^{(tr)}$ instead
of $\rm f_m$ was to be kept constant during variations of $\rm \langle
A_v \rangle $  over a given time-step, it would  correspond to an $\rm
f_m$  that instantaneously follows  the variations  of $\rm  \langle A_v
\rangle  $.  This  may not  be entirely  without merit  since whatever
processes are  responsible for setting  up cloud structures  that obey
power laws like that expressed by  Eq. 8 may be also responsible for a
fast  concurrent H$_2$ formation  \cite{CP87, BO91}.   Nevertheless at
this stage  we chose not to  consider this possibility  since the 
  cloud power laws are  assumed ``frozen'' during our simulations and
allowing   instantaneous effects on $\rm f_m$  (via variations of
$\rm \langle  A_v \rangle $)  would run counter  to the spirit  of our
effort to model the time evolution of the HI, H$_2$ gas phases without
such influences.  In the future  we intend to explore this possibility
to  see whether  it is  important when  compared to  the FUV-regulated
HI/H$_2$ phase interplay explored here.

\subsubsection{The applicability of the H$_2$ formation model at high resolutions}

In order to track dense gas our simulations employ SPH particle masses
of $\rm m_{SPH}=500\  M_{\odot }$ (see section 4.2.1)  where it can be
argued that the applicability of  the scaling laws used to deduce $\rm
f_m$ is  doubtful because  such small masses  are well below  those of
even  the smallest GMCs  and thus  represent cloud  fragments instead.
However,  individual  SPH  particles  do not  represent  the  smallest
resolved objects of  our simulations, rather this is  set by the total
number of particles used to derive the formal SPH estimate of the mean
density.   The latter  is  of  the order  $\rm  M_{ref}\sim \rm  10^4\
M_{\odot }$  (see Section 4.2), comparable to  small molecular clouds.
Furthermore, it  must be noted that the applicability of the  n-R 
scaling law for their $\rm M_{ref}$ sub-units is based  on the virial 
theorem and  the universality of the linewidth-size relation.   The 
latter has  now been confirmed  to hold for scales  and environments 
well inside typical  molecular clouds and with  the same  normalizing 
constant  as for  entire  clouds, a  fact attributed to  the 
universality of large-scale  driving mechanisms for turbulence (Heyer 
\& Brunt 2004).  Regarding the virial theorem, it is a simple 
corollary that if GMCs  are virialized objects so will be any of their
 sub-units albeit at different boundary pressures, thus Eqs 8, 10 are
expected  to hold even for $\rm  M=M_{ref}$ cloud masses making
up larger  GMC-type ensembles ($\sim  (10^5-10^6)\rm M_{\odot}$).  The
latter of course assumes that  the correct boundary pressure $\rm P_e$
is used  in Eqs  8, 10, which  as we  already discussed may  be poorly
represented  by the formal  SPH estimate  of the  macroscopic velocity
dispersion. Thus our model is applicable for scales where the 
size-linewidth relation and the assumption of virial equilibrium are 
valid, which will certainly not be the case in dense star forming cores,
but does seem to be true at the larger scales as probed by galaxy scale
simulations.

\section{An application to dwarf  galaxies}

Our first  application will  be a model  of a dwarf  irregular galaxy.
There  are a  number of  practical reasons  to choose  this type  of a
system as  a test model, as  well as some interesting  issues that are
particular for  dwarf galaxies  that can be  explored using  our model
(e.g. H$_2$ vs HI gas supply).   The small size of these systems allow
relatively high  resolution with modest  computational effort.  Indeed
the  choice of  dwarf galaxies  as modeling  templates allows  a given
numerical  simulation to  probe  small physical  scales  and high  gas
densities,  the  latter  being  of  crucial  importance  if  the  $\rm
HI\rightarrow  H_2$ phase  transition is  to be  tracked successfully.
Insight  gained from  simulating these  systems, whose  properties are
constrained by a wealth of observational data (Van Zee 2001; Barone et
al. 2000; de Paz, Madore  \& Pevunova 2003).  Furthermore it holds the
promise of yielding good constraints about physical processes that are
expected to be universal in  galaxies (see e.g.  similar work done for
SN feedback strength; Pelupessy, van der Werf, \& Icke 2004).

Aside from being excellent testbeds for quantifying phenomena expected
to be common across the  Hubble sequence, dwarf galaxies are important
systems  on  their own  right  because  of  their   possibly  very
  important role in current galaxy formation theories as the building
``blocks'' of larger systems  \cite{KWG93}, and as major ``polluters''
of the intergalactic medium with metals \cite{FT00}.

\subsection{Simulation setup} 

We construct a simple model of a dwarf irregular with gas mass of $\rm
M_{gas}=10^8~   M_{\odot    }$   and   a   stellar    mass   of   $\rm
M_{star}=10^8~M_{\odot}$.  Specifically, the  gas disk has radial 
surface density profile

\begin{equation}
 \rm  \Sigma= \Sigma_g/(1+R/R_{g}),
 \end{equation}

\noindent 
with central density $\rm \Sigma_g=10~M_{\odot}/pc^2$ and radial scale
 $\rm R_g=0.33~kpc$, truncated at 4 kpc. An exponential stellar disk, 
 
\begin{equation}
 \rm 
 \rho_{disk}(R,z) = \frac{\Sigma_0}{2 h_z} \exp(-R/R_d) sech^2(z/h_z)
 \end{equation}

\noindent
with central  surface density $\rm  \Sigma_0=300~M_{\odot}/pc^2$, $\rm
R_d=0.5~kpc$   and  vertical  scale   height  $\rm   h_z=0.2~kpc$,  is
constructed as in Kuijken  \& Dubinski~(1995). The gaseous and stellar
components  are represented  by $2  \times 10^5$  particles  each. For the
gravitational softening length of the stellar particles a value of 
20 pc is adopted,  while for the gas  particles it is taken to be 
equal to the  SPH smoothing length.  The ages of the initial
population of stars are distributed according to a constant
star formation rate (SFR) and  an age of  13 Myr. \\
\noindent 

The rotation curves of dwarf galaxies are generally best fit using 
dark halos with a flat central core (Flores \& Primack 1994,  
Burkert  1995). Therefore we take a halo profile   
 
\begin{equation}
 \rm \rho_{halo}(r)= \rho_0 \frac{\exp(-r^2/r_c^2)}{1+r^2/\gamma^2}
 \end{equation}

\noindent 
with core  radius $\rm  \gamma=2~kpc$, cutoff radius  $\rm r_c=20~kpc$
 and central density $\rm \rho_0=2 \times 10^7~M_{\odot}/kpc^3$, for a
 total mass of $\rm M_{halo} =  15 \times 10^9\, M_{\odot}$ and a peak
 rotation velocity of about 50 km/s. Note that the profile adopted is 
 very similar to the Burkert (1995) profile. They differ mainly in their
 asymptotic behaviour for $\rm r \to \infty$, thus the Burkert profile will
 only deviate significantly well outside the region of interest for our 
 simulation (within 20 kpc the difference in rotation velocity of the 
 adopted density profile and a Burkert profile with the same central 
 density is $<6\%$). We represent the dark-matter halo by a static potential.

\subsection{Practical considerations}

The mass resolution  of the simulation limits the  density that can be
probed. We can derive some minimum requirements in order for our model
to  be able  to  follow  the $\rm  HI\rightarrow  H_2$ transition.  In
addition, it may be necessary to consider the effect of H$_2$ cooling.
 Finally, unlike previous numerical work reported in the 
literature, the range of values for the delay factor $\rm f_{sf}$, which 
regulates star formation, can now be explored under constraints imposed 
by the observationally-motivated demand that H$_2$ forms ahead of stars.

\subsubsection{The necessary resolution: the choice of M$_{ref}$}

Simulations   of  self-gravitating   fluids  done   with  insufficient
numerical resolution can suffer from artifacts: artificial clumping or
inhibition of  gravitational collapse may occur.   For SPH simulations
including self-gravity these effects can arise if the local Jeans mass
is not  resolved (Bate \&  Burkert 1997; Whitworth 1998).  Hence, this
requires for our simulation

\begin{equation}
\rm M_J 
 > N m_{SPH},
\end{equation}

\noindent
with  N  the number  of SPH  neighbours and  $\rm
m_{SPH}$  is the mass  of an  SPH particle.   In our  simulations this
requirement is met by choosing appropriate star formation parameters so
that violation of  Eq. 27 is precluded by  star formation.  A gas
particle  will ``spawn''  star  particles, and  thus  be subjected  to
heating  that  raises  the   local  Jeans  mass,  whenever  $\rm  M_J<
M_{ref}$. Taking $\rm M_{ref}=N m_{SPH}$ in effect makes the choice of
mass  (and  density)  resolution  equivalent  to the  choice  of  $\rm
M_{ref}$.   The density  resolution of  our simulations  must  be high
enough so that H$_2$ formation can compete with FUV dissociation (i.e.
$\rm r_{dis}<  1 $,  Eq. 13).   In Figure 1  it is  apparent that
densities of $\rm \sim (50-100)\ cm^{-3}$ are sufficient to follow the
$\rm  HI\rightarrow H_2$ transition,  while somewhat  higher densities
may be necessary  for low metallicity gas.  For  such densities and at
typical $\rm T_k(CNM)\sim 100\  K$, Eq. 18 yields $\rm M_J\approx
10^4\ M_{\odot}$.   Hence running the simulation  with particle masses
of $\rm  m_{SPH} \approx 500\  M_{\odot}$ allows a good  resolution of
the  aforementioned  mass scale.   Apart  from  properly tracking  the
HI/H$_2$  phase transition,  setting  a small  $\rm M_{ref}\sim  10^4\
M_{\odot  }  $  allows  GMC-type,  star-forming  associations  of  gas
particles to  naturally emerge in the  simulations.  Indeed, GMC-class
gas  masses of  $\rm M_{GMC}\sim  (10^5-10^6)\ M_{\odot  } $  can then
emerge as  assemblies of smaller  star-forming cloud units.   Our star
formation  criterion $\rm  M_J<M_{ref}$, where  $\rm M_{ref}<M_{GMC}$,
ensures  that  such self-gravitating  SPH  particle formations  (which
previous  simulations  show  emerging,  e.g.  Gerritsen  1997)  always
contain gravitationally  unstable regions that form  stars, exactly as
in real GMCs.   In that respect $\rm M_{ref}$ can be  viewed as a mass
``radius''  attached to  individual  SPH particles  which defines  the
Jeans-mass instability, and  which is itself smaller than  the mass of
typical  GMC aggregates,  the  preferred sites  of  star formation  in
galaxies.  If one were to choose $\rm M_{ref} \gg M_{GMC}$ stars could
then form from ``cloud'' mass  scales that are never seen star forming
in nature.  As  a final note on the choice of  $\rm M_{ref}$ it worths
mentioning that  observational studies have  shown that O, B  stars do
not  form  in GMCs  with  masses below  $\sim  10^5\rm  \ M_{\odot}  $
(Elmegreen 1990).   Thus choosing  $\rm M_{ref}\sim 10^4  \ M_{\odot}$
makes  our  simulations capable  of  following  the  evolution of  the
smallest GMCs seen forming stars in galaxies.

\subsubsection{The choice of $\rm f_{sf}$: the role of H$_2$ and star formation}

In our  standard star formation recipe the  delay for star 
formation to occur is set by the free parameter $\rm f_{sf}$;  
our default choice is a fixed value of $\rm f_{sf}=10$.  In the 
past typical values of $\rm f_{sf}\sim 5-10 $ were chosen so that the 
expected star formation rate from the  GMCs present e.g. in the Milky 
Way  would be similar to the one observed.  Dedicated high-resolution 
simulations of individual molecular clouds may  shed some light on 
this issue  (e.g. MacLow et al 1998),  but  for  the  time  being  
the  choice  of  $\rm  f_{sf}$  in galaxy-size simulations like ours 
remains rather apriori.

Nevertheless  the fact that  everywhere in the local Universe  stars 
seem to form  out of molecular rather than atomic gas allows us to 
deduce a lower limit on the value of $\rm f_{sf}$.  Indeed, for  H$_2$
formation to remain always  ahead of star formation it must  be 
$\rm \tau_f(H_2) \le \tau_{sf}$,  which from Eqs 2 and 19 yields,

\begin{equation}
\rm f_{sf}\ga 5.33 \left(\frac{\langle n_1 \rangle}{cm^{-3}}\right)^{-1/2}
\left(\frac{\mu Z}{3.5}\right)^{-1} 
\left[\frac{(1+T_k/100~ K)^2}{(T_k/100~K)^{1/2}}\right].
\end{equation}

\noindent
Since CNM HI is the most  likely precursor phase of the H$_2$ gas, for
typical conditions of $\rm T_k\sim  (100-200)\ K$ and $\rm \langle n_1
\rangle\sim  (10-50)   cm^{-3}$,  the  latter   equation  yields  $\rm
f_{sf}\ga 3-10$.   These values are  roughly similar to  those deduced
using  the constraints  on the  observed  star formation  rate in  the
Galaxy.   Given   the  uncertainties   inherent  in  e.g.    the  $\rm
S_{H}(T_k)$ function, such a rough agreement is noteworthy and here it
is  worth  mentioning   that  $\rm  \tau  _f(H_2)$  is   also  a  good
approximation of  chemical equilibrium timescales  in FUV-illuminated
clouds \cite{HT99}.  Thus the rough agreement of the constraint set by
Eq.  28  with what are  considered reasonable $\rm f_{sf}$  values may
signify an important  role for cloud chemistry in  setting the average
star formation rate. This  is not far-fetched since molecule formation
enables  very different  gas  cooling functions  and thus  drastically
alters  its  thermodynamic  state  allowing  to  eventually  cool  and
fragment  further  (well beyond the resolution limit for galaxy-sized 
simulations) (e.g.  Chi\'eze \& des  For\^ets~1987). For these reasons 
we will also discuss a model for the star formation which depends on 
$\rm f_m$ directly. 

\subsubsection{Cooling by H$_2$}

A detailed calculation of H$_2$ cooling  was done by Le Bourlot et al.
(1999).  At $\rm  T_k \sim (1000 - 8000)K $ H$_2$  is a more efficient
coolant than  the other major coolants  (C and Fe).  Hence  even a low
($\sim 10^{-4}$) remnant abundance of molecular gas in the diffuse WNM
can have an impact on  the cooling at these temperatures.  Indeed some
people  have  included H$_2$  cooling  in  their  cooling curves,  but
without   calculating  the   abundance   of  molecules   \cite{CLC98}.
Admittedly our accuracy in following  the amount of warm molecular gas
is limited, but to get some idea  of the possible effect of such a gas
phase it is nevertheless interesting to do some calculations including
H$_2$ cooling. For this we use  the Le Bourlot cooling curve, where we
use a limited  set of data (low density  limit, constant ortho-to-para
ratio,  only   H$_2$-H  collision  excitation),   appropriate  for  our
purposes.  Note  that some observations  find abundances of  H$_2$ gas
the order of  $10^{-4}-10^{-3}$ in diffuse HI gas,  even in regions of
the ISM hostile to H$_2$ formation, like galactic high velocity clouds
\cite{RS01}.

\section{Dwarf galaxy model: results}

  In  a model  that  incorporates H$_2$  formation  the two  obviously
important parameters  that should be explored are:  the formation rate
$\mu$, and  the metallicity Z.  The  formation rate is  uncertain by a
factor  of   5-10,  and  thus  introduces  some   uncertainty  in  the
theoretically  calculated  H$_2$ content, e.g.  for  high formation  rates
significant amounts  of warm and diffuse  H$_2$ may be present  in the outer
parts of  spiral galaxies (Papadopoulos,  Thi, \& Viti  2002).  Metals
play  an  important  role:   the  H$_2$  formation  rate  is  directly
proportional to the  dust surface available, and we  take this surface
to be proportional to Z.   Dust also plays an important role shielding
molecules from UV radiation.  Observations of low metallicity systems,
like dwarf  irregulars, using  CO as a  tracer give low  molecular gas
contents   \cite{BHH00}.   However,   the   interpretation  of   these
observations  is complicated by  the fact  that the  conversion factor
from CO flux  to H$_2$ mass under these  conditions depends heavily on
metallicity and  ambient FUV  field: for low  metallicity, FUV-intense
environments CO dissociates while  the largely self-shielding H$_2$ is
not affected by much \cite{MB88, I97}.

In this work we explore a 2x2 grid of models using a formation rate of
$\mu=3.5$   and   $\mu=17.5$,  and   metallicities   of  either   $\rm
Z=Z_{\odot}/5$ or  $\rm Z=Z_{\odot}$.  Note  that for our  model dwarf
galaxy the high metallicity  is somewhat unrealistic, because normally
dwarf  galaxies  have  lower   metal  content.   Note  also  that  for
simulations without H$_2$ cooling  our subgrid model follows the H$_2$
content as  a passive tracer of  the history of  the local conditions,
thus the simulations for high and low formation rate will be identical
except for the amount of H$_2$.

\subsection{Star formation properties of the simulations}

We  start the  simulation  with an  isothermal  gas disk  at $\rm  T_k
\approx  8000~K$, and  a molecular  fraction $\rm  f_m=0$.   After the
start of  the simulation the gas  cools and collapses   along the
z-direction, forming  regions of  cold and dense  gas in  the midplane
(the minimum size of structures resolved in the simulation is about 30
pc).  There,  H$_2$ starts  to form and  star formation  starts taking
place as well.  The fact  that we start from an unrealistically smooth
initial  condition  results  in  some  transient  effects,  but  after
approximately 200 Myr the galaxies  settle in a stationary state, with
a constant  star formation (see  Figure \ref{fg_sf}). For  this model 
galaxy,   the  mean star formation rate is   about  $\rm SFR=0.002~M_{\odot}/yr$ for low Z and 
$\rm SFR=0.003~M_{\odot}/yr$ for high Z,  with modest variations  
(the gas depletion timescale  is much longer than the duration of our 
simulation, in the order of $5 \times 10^{10}$~yr).   The  star 
formation  values  we  get represent  fairly typical values for a low 
surface brightness dwarf \cite{Z01}. Also plotted  in  Figure 4  is 
the average  star  formation  density as  a function  of  radius, 
note   the  exponential  distribution  of  star formation and the sudden
truncation. During the simulation a small fraction ($1 \times 10^{-3}$) 
of the gas is expelled out of the disk,  but most of this falls back. A 
hot galactic wind is not formed, the low mass loss rate associated with 
the expected wind for this model cannot be resolved by our current 
mass resolution.

\subsection{Spatial and temporal distribution of H$_2$ gas}

An equilibrium fraction of H$_2$ is reached on timescales of 
$\approx 50-100$  Myr (see  Figure \ref{fg_fmtime}), in  agreement with
  our earlier approximate  estimates from Eq.  2.  
Furthermore,  as expected, the  low-$\mu$/low-Z simulation  yields the
lowest  equilibrium molecular  fraction of  $\rm f_m  \sim  0.001$ and a
higher  formation rate  $\mu =17.5$  will boost  that to  $\rm f_m\sim
0.03$.   For high  metallicities, molecular  fractions of  $\rm f_m\sim
0.17$ ($\mu=3.5$) and $\rm f_m\sim 0.4$ ($\mu=17.5$) form.  Also drawn
in  Figure \ref{fg_fmtime}  is the  time dependence  of  the molecular
fraction for a low-$\mu$/low-Z simulation with logotropic clouds.  The
difference between constant density  and logotropic clouds is not very
large, certainly smaller than the differences due to the uncertainties
of $\mu  $ (this is  also evident in  the equilibrium values  of $\rm
f_m$  shown in  the  panels in  Figure  1, where  the main  difference
between constant density and logotropic clouds was that for logotropic
clouds  appreciable   fractions  $\rm   f_m$  appear  also   at  lower
densities).

In Figures \ref{fg_rfm} and  \ref{fg_hfm} we show the (mean) molecular
fraction as a  function of radius and height  above the disk-plane for
the four simulations. We see  that, as expected, H$_2$ forms mostly in
the central  regions and  in the midplane  where the pressure  and gas
density  builds up.   Note also  that  for high  metallicity and  high
formation  rate the  molecular fraction  seems to  have a  tendency to
''saturate'' in  the central  regions at a  fixed value, in  this case
$\rm f_m=0.6$.  Looking in more  detail at the spatial distribution of
HI and H$_2$ shown in Figure \ref{fg_hih2maps} we find H$_2$ mainly 
concentrated in the dense  clumps and filaments of the HI distribution. 
 The structure of the ISM can be quantitatively compared with 
observations by analyzing the power spectrum of the HI maps such as 
those in Figure \ref{fg_hih2maps}. This is done in Figure \ref{fg_pw}, 
where we see that it agrees quite well with power spectra made for 
the most detailed studied dwarf galaxy, the Large Magellanic 
Cloud (LMC), and is stable over the course of the simulation and as a 
function of resolution. If we now compare the  low  and high $\mu$ 
simulations, we see similar H$_2$ distributions: {\it the uncertainty 
in  the formation parameter is mainly an  uncertainty in the amount of 
H$_2$  formed, not in the locations of  H$_2$ formation}. The relative
distribution of H$_2$ is well represented  by our simulation, but the 
exact amount of H$_2$ is more difficult to determine,  as this depends
on the formation rate parameter.  If we look at the high metallicity 
simulations, we can see that, apart from the clumpy  distribution 
there is also a sizable mass of ''diffuse'' H$_2$, but still confined 
mainly in gas filaments.

Finally  we look  at the  relation  between star  formation and  H$_2$
distribution.   In  Figure   \ref{fg_h2hamaps}  we  show  a  simulated
''H$\alpha$''  map  (actually  a  map  of the  stellar  luminosity  in
ionizing photons),  overplotted with contours of H$_2$  gas, chosen so
that  they highlight  the relation  of star  formation to  the densest
parts of the  H$_2$ distribution.  It can be  seen that star formation
is almost always  associated with some nearby H$_2$  gas complex while
some  of them,  being without  newly formed  stars at  that particular
instant, do not show any H$\alpha$ emission.

\subsection{The influence of the collapse delay factor $\rm f_{sf}$}

Apart from  the parameters directly  related to H$_2$  formation, $\rm
f_m$ will also  depend on the delay factor  $\rm f_{sf}$. As discussed
in section 4.2.2,  this parameter accounts for the  fact that a region
that  is  Jeans-unstable  does  not  collapse to  form  stars  at  the
free-fall  time  $\rm  \tau_{ff}$.   The  total  amount  of  H$_2$  is
sensitive to the time HI gas stays in a cold and dense phase conducive
to H$_2$ formation, while the emergence of new stars from such a phase
helps  dissociate H$_2$  by  dramatically increasing  the ambient  FUV
radiation.   Indeed, comparing  models run  with different  values for
$\rm f_{sf}$ reveals the total  amount of molecular gas to be strongly
dependent on this  parameter (Figure \ref{fg_fmtime_tcol}).  In short,
the formation of stars and  H$_2$ are in competition, and small enough
delay factors can  quench H$_2$ formation.  Yet, a value of
$\rm f_{sf} \gtrsim 10$, as  constrained by the observed GMC  masses 
and the star formation rate in the Galaxy, amply satisfy Eq. 28 and 
H$_2$ formation preceeds star formation.

\subsection{Star formation with an H$_2$ gas fraction threshold}

In section 4.2.2 we have argued for a connection between star formation 
and chemical timescales, the latter well approximated by the H$_2$ formation 
timescale. A simple and direct implementation of this idea is to replace 
the preset delay time $\rm \tau_{sf}$ by a threshold molecular fraction 
$\rm f_{m,sf}$ for the onset of star formation. In this case the delay 
factor $\rm f_{sf}$ is no longer needed; the ``delay'' from a pure free-fall 
timescale is now the outcome of the interplay between the competing physical 
processes that lie behind the establishment of $\rm f_m$. 

We rerun the $\rm Z=Z_\odot, \mu=3.5$ simulation with this new star 
formation recipe and a threshold value of $\rm f_{m,sf}=1/8$. In figure 
\ref{fg_fmtime_tcol} we have plotted $\rm f_m$ for this case. As can 
be seen, the equilibrium molecular fraction goes down significantly, 
to $\rm f_m \approx 0.01$, which is easily understood, 
because such a low threshold value will almost immediately destroy H$_2$ 
once it forms. For higher threshold values $\rm f_m$ is higher, 
$\rm f_m \approx 0.05$ and $0.1$ for $\rm f_{m,sf}=3/8$ and $5/8$ respectively 
(this is still less than for the normal recipe, because for this case about $40\%$ of
H$_2$ is in regions with $\rm f_m > 5/8$). The star formation rate for these 
 models is about $\rm SFR=0.004~M_\odot/yr$.

In figure \ref{fg_tausf} we plot the resulting (cumulative) 
distribution  of delay times $\rm \tau_{sf}$, normalized on the 
local free-fall time $\rm t_{ff}$. A higher threshold results in longer 
average delay times, from $\rm \langle \tau_{sf} \rangle \approx 1.5 t_{ff}$ for 
$\rm f_{m,sf}=1/8$ to $\rm \langle \tau_{sf} \rangle \approx 10 t_{ff}$ for 
$\rm f_{m,sf}=5/8$. However, the fraction with $\rm \tau_{sf} < t_{ff}$ 
stays roughly constant. The reason for this is that once star formation starts
in a region, it takes some time for feedback to kick in. In the mean time 
molecule formation continues, and the molecular regulated star formation 
allows for extra star formation events if enough H$_2$ forms (in the 
previous implementation of star formation there was no such possibility). 
In other words, \emph {the local star formation efficiency is 
partly determined by the H$_2$ formation.}

A high value for $\rm f_{m,sf}$ seems to be favoured, because star formation
regions are observed to be mainly molecular, and in this case both the
H$_2$ fractions and local star formation rates are consistent with what is known 
from observations. Hence this model for star formation seems promising for 
future application.

\subsection{Simulations with H$_2$ cooling}

The inclusion of  H$_2$ cooling affects the simulation  in a number of
ways. As an extra coolant it will increase the amount of gas in a cold
state, and in turn this will increase the formation of H$_2$. However,
star formation  will also  increase, and thus  also the UV  field.  In
Figure  \ref{fg_fmcool}  the  effect   of  cooling  on  $\rm  f_m$  is
illustrated for  the $\mu=3.5$ simulations.   As can be  seen, cooling
has some influence for the  high Z simulation but for low Z  the
molecular fractions  are too low  to substantially alter  the results.
Note that while H$_2$ is a  strong coolant at high temperatures (a few
thousand K), where  it dominates the cooling for  $\rm f_m>0.001$, the
effect of the extra cooling is  limited because of the small amount of
gas at these temperatures.

\section{Discussion}

Detection and mapping of the  molecular component of dwarf galaxies is
usually   done   by   mapping   the   $\rm   ^{12}CO   (J=1-0)$   line
\cite{BHH00,MY01}   or  by   UV  absorption   studies   (Tumlinson  et
al. 2002). The derived H$_2$ fractions  are $ \sim (1-10)\% $, but for
low  metallicity systems  CO is  often not  detected  corresponding to
upper limits  for the  molecular fraction of  $\sim 1\%$.   Recently a
large observational  effort has detected $ ^{12}$CO  J=1--0 in several
more  dwarf  galaxies  increasing  the  number of  such  systems  with
detected CO emission by $\sim 50\%$ (Leroy et al. 2005).

 At this stage  a direct comparison of our  simulations with available
CO  imaging  data  must  be  made cautiously.   This  is  because  our
simulations follow  the formation of H$_2$  not CO. The  latter is the
most  abundant molecule  after H$_2$  itself and  serves as  the prime
tracer of  its mass (e.g. Dickman,  Snell, \& Schloerb  1986; Young \&
Scoville 1991), yet it is still four orders of magnitude less abundant
and  thus, unlike  H$_2$, it  cannot  self-shield.  This  is the  main
reason  why  in metal-poor  and  FUV-intense  environments like  those
prevailing in  dwarf galaxies CO  (but not H$_2$) will  be dissociated
and  its emission can  then give  a rather  misleading picture  of the
H$_2$  mass distribution  (e.g.   Madden  et al.   1997).   Even if  a
simplified version  of the  chemical network giving  rise to  CO (e.g.
Nelson  \&  Langer 1997)  were  to  be  included in  the  simulations,
densities $\rm  n\ga 500\ cm^{-3}$  and temperatures $\rm T_k<50\  K $
would have to be tracked  in order to successfully model CO formation,
and these  are currently inaccessible by  galaxy-size simulations like
ours.

In  principle   one  could  use   empirical  relations  of   the  $\rm
X=M(H_2)/L_{co(1-0)}=F(G_{\circ},  Z, n,  T_k)$ factor  available from
the literature (e.g.  Israel 1997;  Bryant \& Scoville 1996), and then
convert the  H$_2$ maps  from our simulations  to CO  brightness maps.
This we  intent to  do in  future work that  will also  include spiral
disks.  Nonetheless, the molecular fractions we derive here seem to be
reasonable, although  for the low-$\mu$/low-Z  simulations the derived
$\rm f_m$  is on  the low side.   The molecular fraction  may increase
somewhat  for  a  simulation  run  at higher  resolution,  because  in
metal-poor  environments the  $\rm HI  \to H_2$  transition  is taking
place  above $\rm  n=100~cm^{-3}$, this  can increase  $\rm f_m$  by a
factor 2-5.   For the galaxy  models run with high  metallicity and/or
formation rate we  find quite substantial fractions of  H$_2$, of $\rm
f_m=0.17-0.4$.

We  also  seem  to  find  quite  high  values,  $\rm  f_{H_2}  \approx
0.001-0.01$, of the  diffuse warm neutral medium to  be molecular.  To
illustrate  this  we  have   plotted  in  Figure  \ref{fg_hivsh2}  the
molecular fraction as a function of total column density for a face-on
projection.   There  is  a  clear  column density  threshold  for  the
presence  of  H$_2$ for  the  low  metallicity  simulation.  For  high
metallicity even  low column  density regions (which  are also  of low
density  and high temperature)  show fractions  of $\rm  f_{m} \approx
0.001-0.01$. This state of the  gas (low density, high temperature) is
not  typically associated  with  the presence  of  such quantities  of
H$_2$.  There  is no  - or  very little -  formation of  molecular gas
going  on  in  this  gas  phase  and  any  $\rm  H_2$  present  should
dissociate.  In  fact, destruction is taking place  in the simulation,
but its timescales  are quite long. For the  thermal dissociation this
is  due to  the  low density  of  the gas:  for  temperatures of  $\rm
T=5000~K$  and density  of  $ \rm  n_H=0.3\  cm^{-3}$ the  collisional
destruction  timescale (for  neutral gas)  is $\rm  \tau_{col} \approx
10^8~yr$.  Furthermore,  for our model galaxy the  radiation fields in
quiescent regions  are quite low, $\rm  G_0 \lesssim 0.1$,  due to the
low star formation, so  the radiative dissociation timescales are also
in the  order  of $\rm \tau_{rad}=1.3/G_0  \times 10^7~yr \approx
10^8~yr$.   Hence the  H$_2$  survival in  lukewarm gas  may not  be
unrealistic.  Note  however that the simulation may  not represent the
dispersion of H$_2$ from dense regions realistically: molecular gas is
formed  in dense,  star forming  regions, where  radiation  fields are
higher  (for our model  up to  $\rm G_0=100$.   As the  star formation
heats  the  surrounding gas  it  will  destroy  H$_2$, the  amount  of
molecular gas  that ''escapes'' destruction is then  sensitive to e.g.
the time it is exposed to  the high radiation field or the combination
of  high density/  high temperature.   Also, destruction  by supernova
shock is  not represented  very well, because  the interaction  of the
supernova shocks with the (sub resolution) molecular clouds is absent.
On the other hand shocks have  been shown recently to also induce fast
H$_2$ formation  as they propagate  through diffuse HI gas  (Bergin et
al. 2004)  and the effect of  such a mechanism in  the overall HI/H$_2$
balance in our simulations is unknown.

Nevertheless,  our simulations do  indicate that   in so  far as  the
molecular  gas  is  not destroyed at  the  sites  of  its  original 
assembly,  it can  survive the CNM-to-WNM transition. In that case 
  the  lukewarm  and WNM  gas  phases  may  still contain  significant
  amounts  of H$_2$.   In subsequent  work we  intend to  explore the
issue of H$_2$ survivability  under such conditions more thoroughly by
introducing  a  better  model   of  its  collisional  and  FUV-induced
destruction  in such  gas  phases while  also including  shock-induced
H$_2$ formation and destruction mechanisms. Finally, any significant 
underlying cloud substructure in CNM HI clouds is likely to enhance 
the amount of H$_2$ gas present in that phase.
  
Future applications of H$_2$-tracking numerical models of galaxies can
be  divided in  two  broad and  complementary  categories, namely:  a)
examine  the effects  of the  various  model parameters  on the  H$_2$
distribution  within a  given galaxy  by e.g.   using  different H$_2$
formation functions (see Cazaux \& Tielens 2004), and/or using a range
of $\rm  f_{sf}$, $\rm  \epsilon _{SF}$ and  $\rm M_{ref}$  values, b)
explore the  evolution of the H$_2$/HI distribution  across the Hubble
sequence using a single  ``mean'' H$_2$-tracking model.  In the latter
category  starbursts/mergers,  with  their extreme  and  fast-evolving
pressure  and  radiation  environments  (and hence  probably  strongly
varying  H$_2$/HI  fractions)   provide  particularly  attractive
templates for modeling.

\section{Conclusions}

 We  presented  a method  to  calculate  the  local H$_2$  content  in
simulations of the ISM based on  a subgrid model for the formation and
destruction of molecular gas.  The model tracks the formation of H$_2$
on dust grains and its destruction  by UV irradiation in the CNM phase
and its collisional destruction in the WNM phase including the effects
of shielding  by dust and  H$_2$ self-shielding.  The solution  to the
H$_2$  formation/destruction  problem  is  simplified greatly  by  the
assumption that H$_2$ formation takes place in structures that conform
to the observed density/size  scaling relations, present on scales not
resolved by the  hydrodynamic code of the simulation.   We show steady
state solutions of the model for the molecular fraction $\rm f_m$ as a
function of the average gas density, temperature, velocity dispersion,
radiation  field,   and  metallicity.    The  effects  of   the  cloud
substructure are explored by taking $\rm 1/r$ density profiles for the
model clouds,  this seem to have  only a modest effect  on the derived
H$_2$ fractions, but it does allow more diffuse gas to be molecular.

A simple time-dependent formulation of  the same model is then coupled
to hydrodynamic simulations and  used to calculate the local evolution
of  the molecular  gas  fraction according  to  the local  macroscopic
quantities.  We have then incorporated our H$_2$-tracking method in an
N-body/SPH code for the simulation of galaxy-sized objects.  Our first
application is to model a  typical low surface brightness dwarf galaxy
where the demands set by  the density resolution necessary to properly
track  the $\rm  HI\rightarrow H_2$  transition are  easily  met.  Our
results can be summarized as follows:

\begin{itemize}

\item  We  found  that   our  model  reproduces  reasonable  molecular
fractions  ranging  from  $\rm  f_m=0.001-0.03$  for  low  metallicity
to $\rm f_m=0.17-0.4$ for solar metallicity systems,  saturating to
$\rm f_m=0.6$ for high values of the H$_2$ formation rate and in the
central regions of our model galaxies.

\item The  biggest uncertainty is  the value of the  adopted formation
rate  parameter $\mu$,  however the  general features  of  the spatial
H$_2$ distribution do not depend significantly on $\mu$.

\item  The  molecular  fraction  shows  a  strong  dependence  on  the
  metallicity Z,  as expected for  H$_2$ forming on dust  grains.  Our
  model  for a  high  metallicity dwarf  shows  surprising amounts  of
  ''diffuse'' H$_2$, possibly due  to the low radiation fields outside
  star forming regions.  This may  be analogous to what happens in the
  outer regions  of spiral galaxies  where the radiation  field drops,
  and a diffuse H$_2$ gas phase may be present there. However a better
  model for  the H$_2$ formation and  destruction in the  WNM phase is
  needed before the aforementioned results can be considered secure.

\item The  molecular gas content is  sensitive to the details of our 
star formation  recipe,  especially on the value of the delay 
factor $\rm f_{sf}$ adopted. This poorly constrained factor, generally 
used to parameterize a slower-than-free fall star formation timescale, needs 
to be sufficiently large, $\rm f_{sf} \gtrsim 10$ to allow for H$_2$ formation 
before young stars start to dissociate H$_2$. On the other hand 
a star formation criterion based on the formation of H$_2$ can dispense
with the $\rm f_{sf}$ factor completely, giving local star formation 
rates well within constraints set by observations. This suggests that 
the star formation is governed by cloud chemistry processes, and indicates 
the importance of considering molecular gas in simulations.

\item H$_2$ cooling has only a minor impact for our model. The largest 
effect of H$_2$ cooling is seen in high metalicity (and thus H$_2$ 
content) environments.

\end{itemize}

Summarizing, we  developed an algorithm to calculate  the evolution of
 the  molecular gas  phase during  the  evolution of  the gaseous  and
 stellar  content of  galaxies, and  obtained a  first  application to
 dwarf  galaxies. Strong  points  are that  it  is simple,  physically
 motivated and  not computationally expensive.   It is well  suited to
 galaxy-sized simulations  where a  complete treatment of  the physics
 involved in the formation of  the H$_2$ clouds is inaccessible, while
 our approach can be seen as complementary of (much higher resolution)
 simulations  modeling  individual   molecular  clouds.   The  set  of
 astrophysical   questions   that   can   be  addressed   using   such
 H$_2$-tracking  models   is  large.   These   include  examining  the
 influence  of ``micro''-physics  parameters/functions like  the H$_2$
 formation  rate on the  general molecular  gas distribution  within a
 given  galaxy, as  well as  modeling  the gaseous  ISM evolution  for
 galaxies across the  Hubble sequence.  The latter can  now be done by
 including all  gas phases  along with the  stars, with  the molecular
 phase  expected to  be the  most  intimately involved  with the  star
 formation.   Finally,  mergers/starbursts,  with their  fast-evolving
 pressure and  radiation environments  are ideal templates  for future
 applications of the model presented~here.

\newpage

\appendix

\section{Analytical approximations for $\rm A^{(tr)}$}

It is  interesting to examine approximate analytical  solutions of Eq.
12, if  only for the reason  of checking the more  general solution in
limiting cases.  This  can be done for two  particular domains, namely
that of rapidly increasing and rapidly decreasing $\rm A^{(tr)}_{FUV}(
t )$.  For $\rm \tau_f dA^{(tr)}  _{FUV}/d t \ll 0$ we can approximate
Eq. 12 as

\begin{equation}
\rm \frac{dA^{(tr)} _{FUV}}{d t}=-\frac{1}{\tau_f} A^{(tr)} _{FUV}\Rightarrow
A^{(tr)} _{FUV}=A^{(tr)} _{FUV}(0) e^{-t/\tau_f}.
\end{equation}

\noindent
In the  case of $\rm  \tau_f dA^{(tr)} _{FUV}/dt  \gg 1 $, and  thus a
decreasing molecular mass fraction, we approximate Eq.  12 as

\begin{equation}
\rm  \frac{dA^{(tr)}_{FUV}}{dt}= \frac{r_{dis}}{\tau_f} e^{-A^{(tr)} _{FUV}}=
2 G_{\circ} k_{\circ} \Phi e^{-A^{(tr)} _{FUV}},
\end{equation}

\noindent
hence

\begin{equation}
\rm A^{(tr)} _{FUV} (t) = ln \left( e^{A^{(tr)} _{FUV}
 (0)}+ 2 G_{\circ} k_{\circ} \Phi t \right),
\end{equation}

\noindent
which, for  $\rm R_f=0$, provides a  convenient solution describing
the photo-destruction of molecular clouds.

\section{logotropic clouds}

Here we give the corresponding  equations of section 2 for clouds with
a density  profile $\rm n(r)=n_e  (r/R)^{-1}$ (``logotropic clouds'').
For a  radially incident interstellar radiation field,  such a density
profile yields

\begin{equation}
\rm N_{tr}(HI)= \frac{\nu _{\circ}}{\sigma} 
ln\left(1+ \nu ^{-1} _{\circ} \frac{G_{\circ} k_{\circ}}{n_eR_f} \Phi\right),
\end{equation}

\noindent
where    $\rm    \nu    _{\circ}=n_e    R   \sigma    \left(1+n_e    R
  \sigma\right)^{-1}$.  For $\rm R\rightarrow  \infty $ we obtain $\nu
_{\circ}\rightarrow  1$ and the  transition column  density of  Eq. B1
reduces that of  Eq.  4 (for $\rm n(r)\rightarrow  n_e$), as expected,
because for  large clouds the density  will not change  much over $\rm
N_{tr}(HI)$.   The  visual   extinction  corresponding  to  this  $\rm
N_{tr}(HI)$ is given by

\begin{equation}
\rm A_v ^{(tr)}=1.086\nu _{\circ}\xi ^{-1} _{FUV}
\ln\left[1+\frac{\nu ^{-1} _{\circ} G_{\circ}}{\mu
 S_{H}(T_k)}\left(\frac{\xi _{FUV}}{Z T_k}\right)^{1/2} 
\left(\frac{n_e}{135\ cm^{-3}}\right)^{-1}\right].
\end{equation}

\noindent

and the H$_2$ fraction $\rm f_m$, 

\begin{equation}
\rm f_m \equiv \frac{M(H_2)}{M_c}=exp\left[-4\frac{A_v ^{(tr)}}{\langle A_v
 \rangle}\right].
\end{equation}

\noindent

The modification of Eq. 12 for the time dependent transition column in
the case  of a  logotropic density profile  (Eq.  22 of  Goldshmidt \&
Sternberg 1995) turns out to be particularly simple, namely

\begin{equation}
\rm \frac{1}{2} \frac{dN(HI,t)}{dt} = \frac{G_{\circ} k_{\circ}\Phi}{\sigma}
e^{-\sigma N(HI,t)} - R_f \int _0 ^{r_{HI}} n(r) n(HI) dr.
\end{equation}

\noindent
Here $\rm r_{HI}$ marks the depth  of the HI layer.  Using the relation
$\rm n(r)=n_e e^{N(r)/N_{\circ}}$ ($\rm  N_{\circ}=n_e R$) valid for a
logotropic density profile, Eq. B4 eventually yields

\begin{equation}
\rm \tau_f \frac{d \sigma N_{tr}(HI,t)}{dt}=r_{dis} e^{-\sigma
N_{tr}(HI,t)}-\sigma N_{\circ} \left(e^{\sigma N_{tr}(HI,t)/\sigma N_{\circ}}-1\right),
\end{equation}

\noindent
where $\rm \tau_f$  and $\rm r_{dis}$ are calculated  for $\rm n=n_e$.
It can be easily seen that for $\rm N_{\circ}=n_e R \rightarrow~\infty$
B5 reverts  back to Eq.  12 with a density $\rm n=n_e$ (i.e.  the case
of a cloud with so large a radius that   the   HI/H$_2$  equilibrium 
is   well   approximated  by   a plane-parallel geometry at a uniform 
density $\rm n=n_e$).

For $\rm  dA^{(tr)}_{FUV}(t)/dt\gg 0$ the  solution of Eq.  B5 remains
the same  as in  Eq. 12  but with $\rm  n_e =2/3  \langle n  \rangle $
replacing n.  Hence, in the case of cloud destruction and $\rm R_f=0$,
Eq. A3 expresses  the time dependence of the  transition layer also in
logotropic clouds.  For $\rm  dA^{(tr)}_{FUV}(t)/dt \ll 0$, Eq.  B5 is
now approximated by

\begin{equation}
\rm \tau_f \frac{dA^{(tr)}_{FUV}(t)}{dt}=-A_\circ
\left[e^{A^{(tr)} _{FUV}(t)/A_\circ}-1\right],
\end{equation}

\noindent
 ($\rm A_\circ=\sigma N_\circ$), with the solution of the latter being

\begin{equation}
\rm A^{(tr)}_{FUV}(t)= 
A_\circ ln \left[ 1+ ( e^{A^{(tr)}_{FUV}(0)/A_\circ}-1 ) e^{-t/\tau_f} \right].
\end{equation}

\noindent
The  latter  expression  reduces  to  that in  A1  when  $\rm  N_\circ
\rightarrow \infty$ ($\rm A_\circ \rightarrow \infty$), as expected.

\newpage

\begin{figure}
\centering
\epsscale{0.465}
\plotone{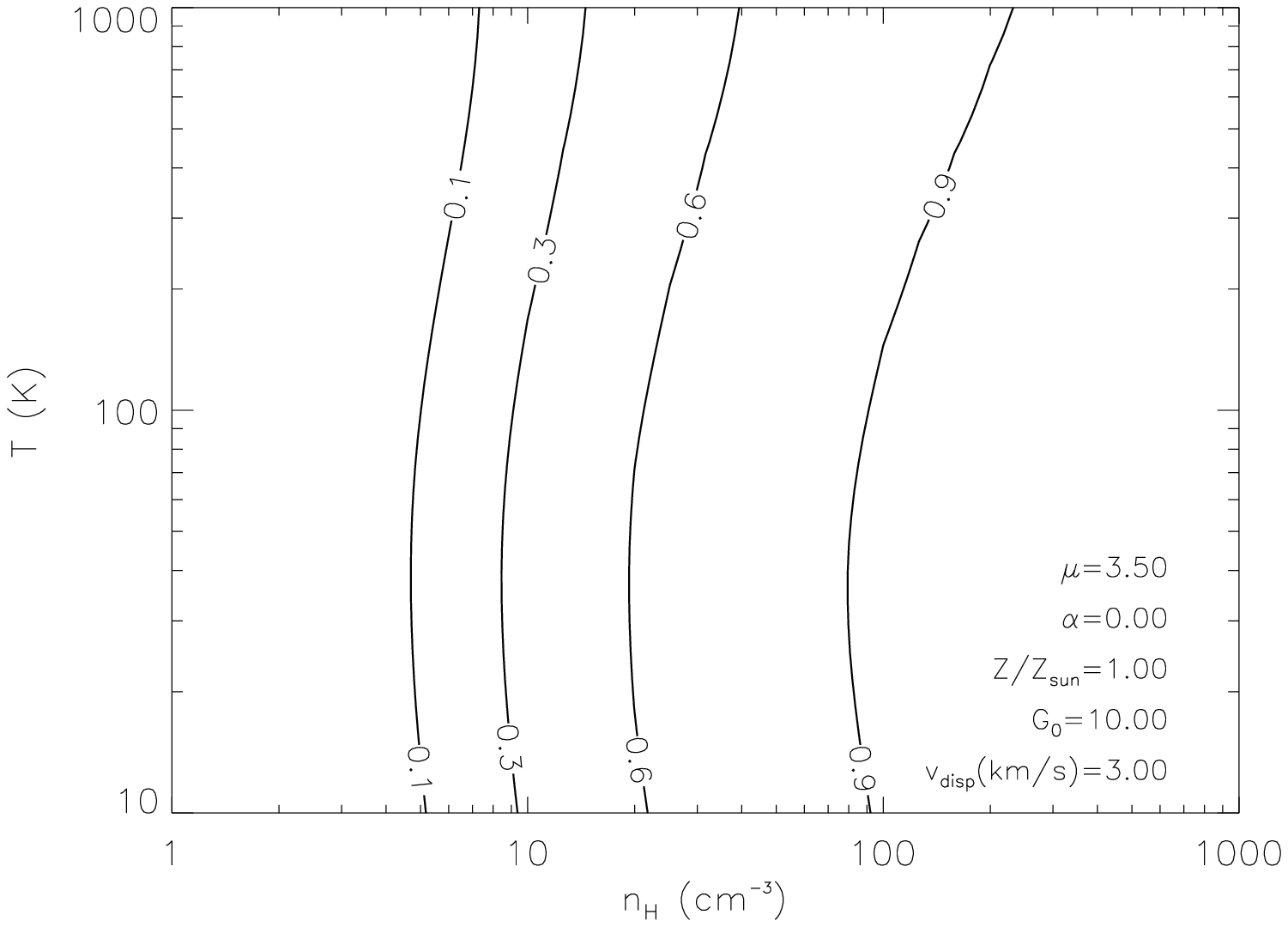}
\plotone{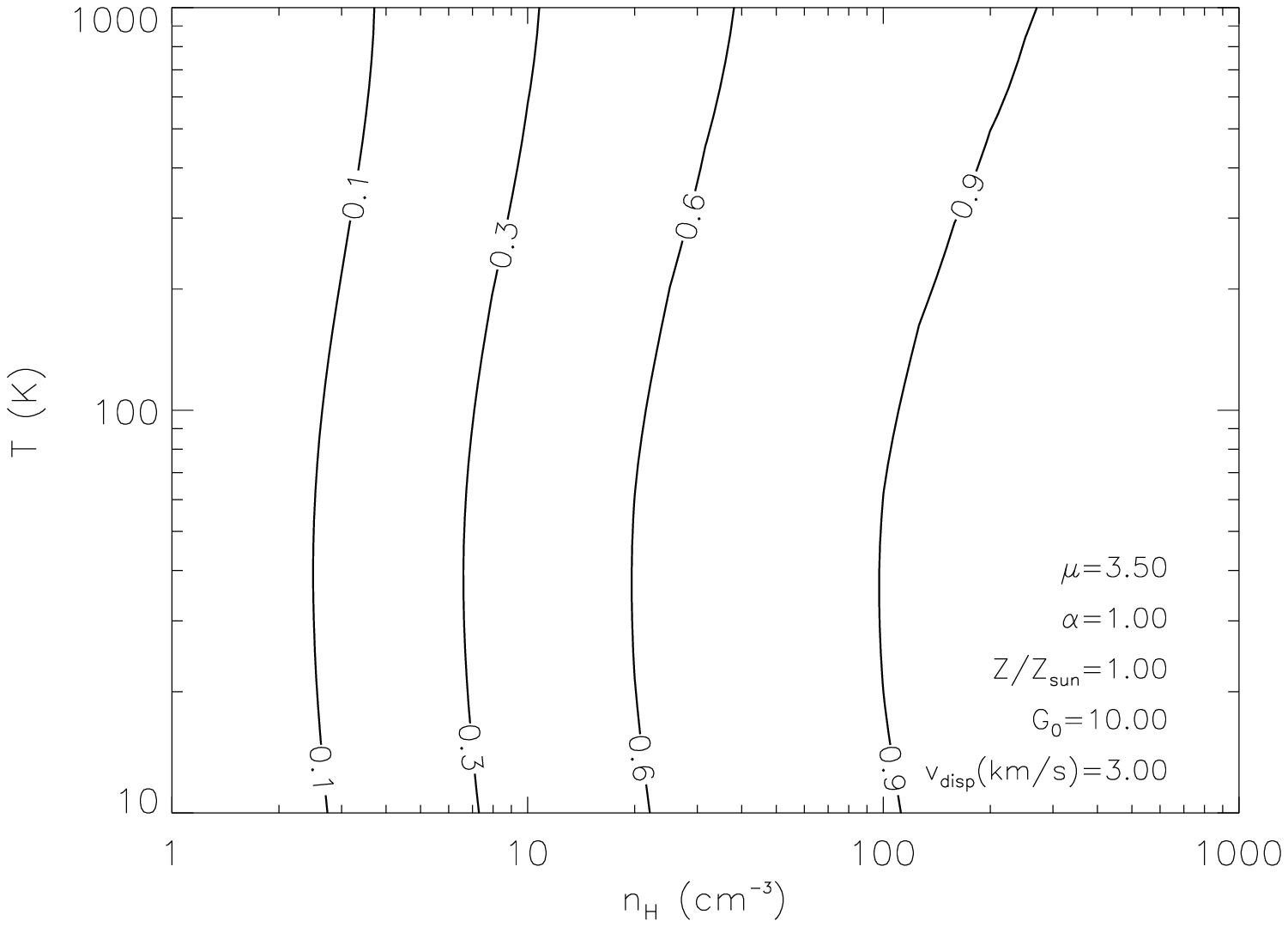}
\plotone{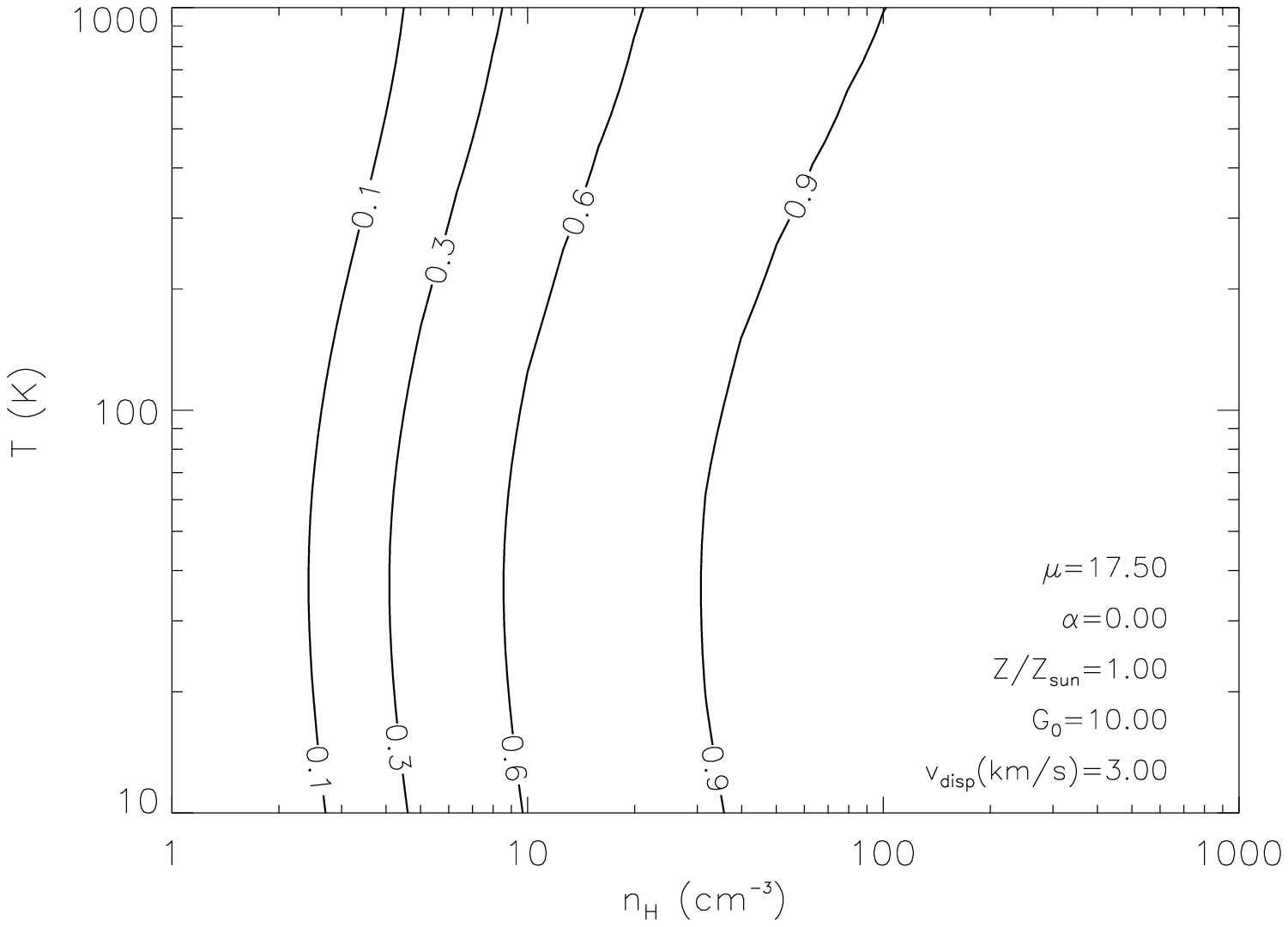}
\plotone{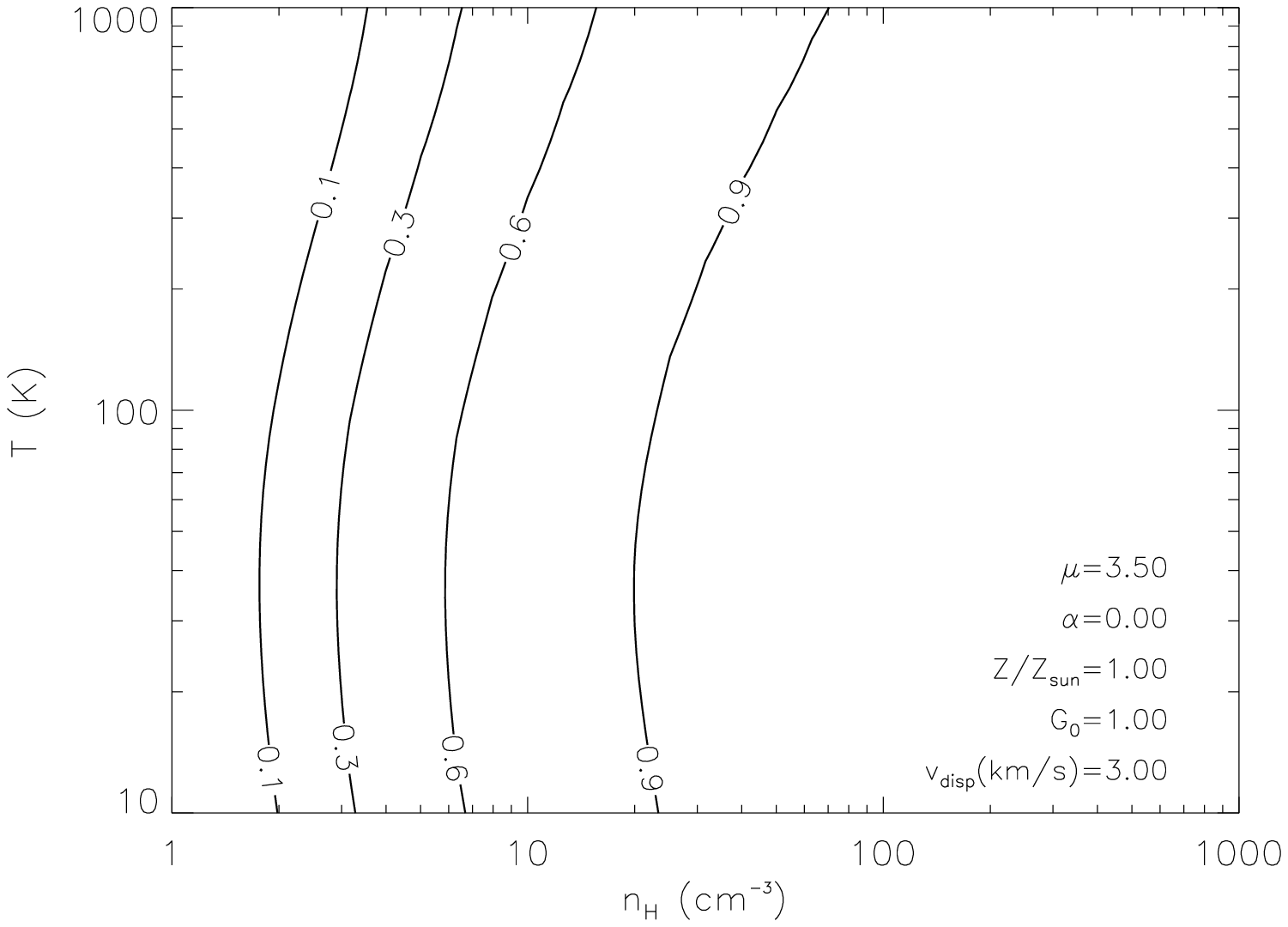}
\plotone{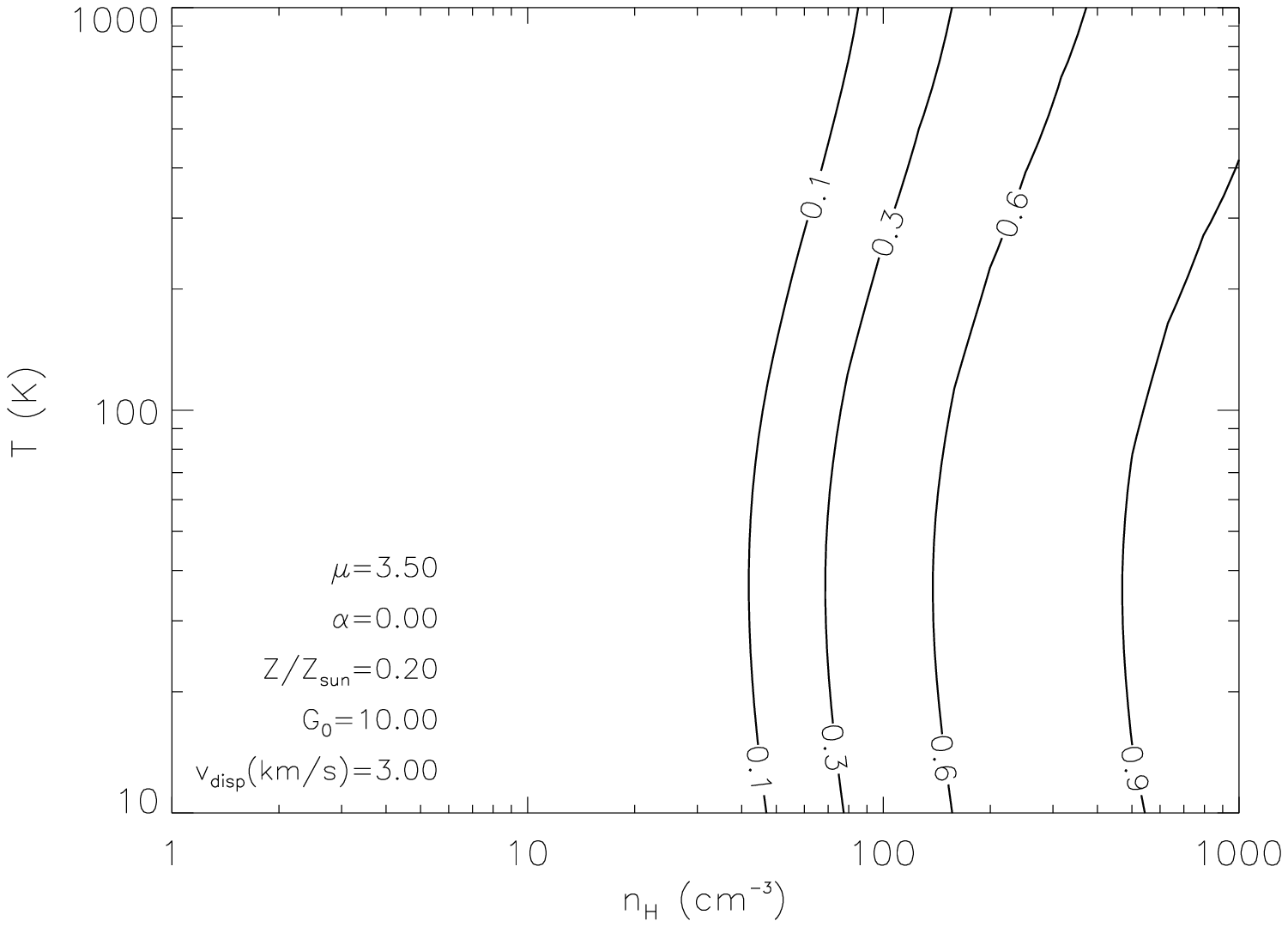}
\plotone{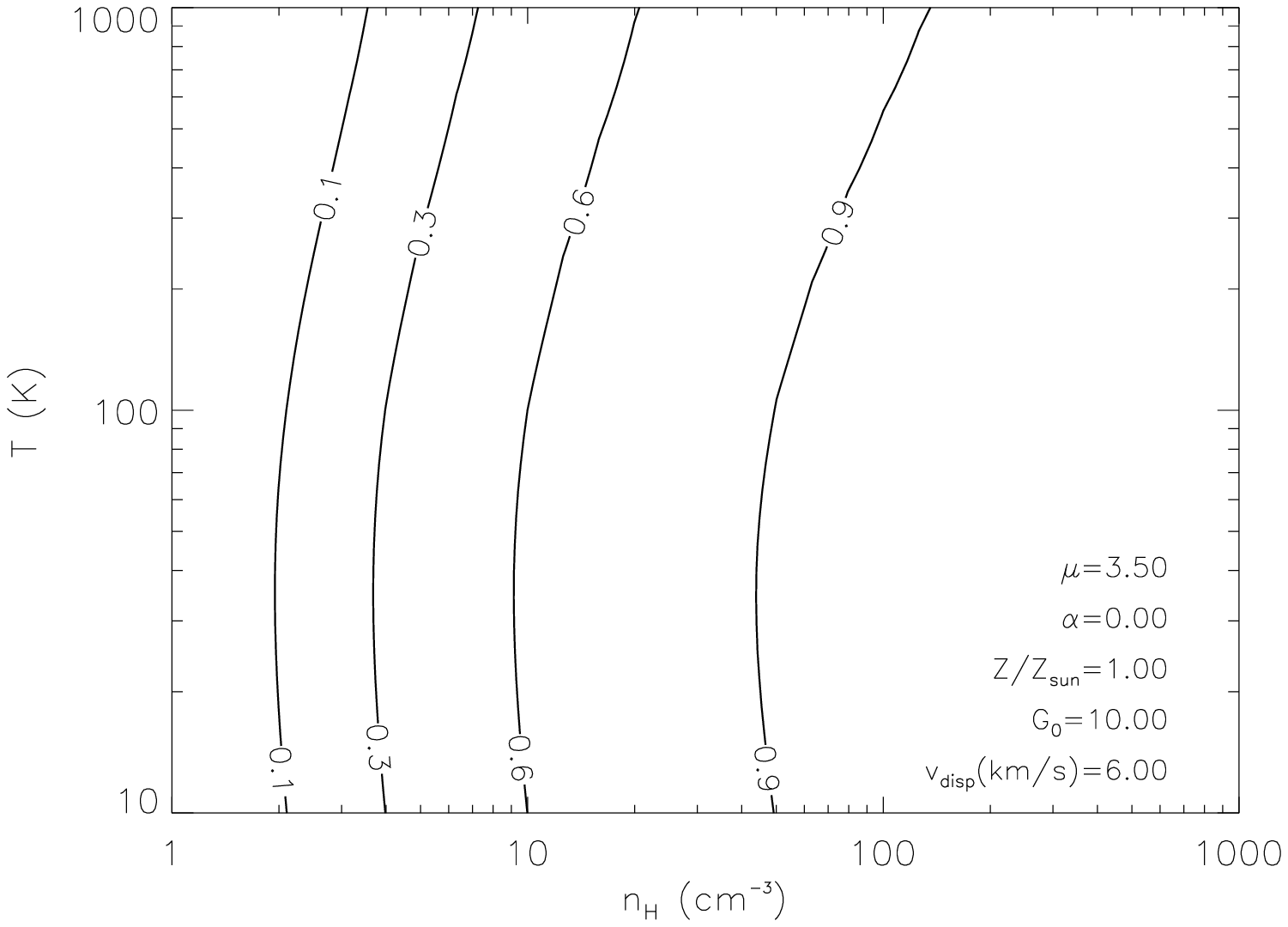}
\caption{The equilibrium  molecular fraction $\rm f_m$.  Plotted
  are contours  (labelled with the  corresponding value) of  $\rm f_m$
  for various representative  combinations of formation rate parameter
  $\mu$,  cloud power  law  index $\alpha$,  metallicity Z,  radiation
  field  $\rm  G_0$ and  velocity  dispersion  $\rm v_{disp}$  (Actual
  values are indicated in the  plots). Above 1000 K the formation rate
  is taken to  be zero. $\alpha$ indicates the power  law index of the
  density profile ($\rm n \propto r^{-\alpha}$), thus $\alpha=0$ means
  constant density clouds, $\alpha=1$ logotropic clouds.}
\label{fg_fh2eq}
\end{figure}

\clearpage

\newpage

\begin{figure}
\centering
\epsscale{.8}
\plotone{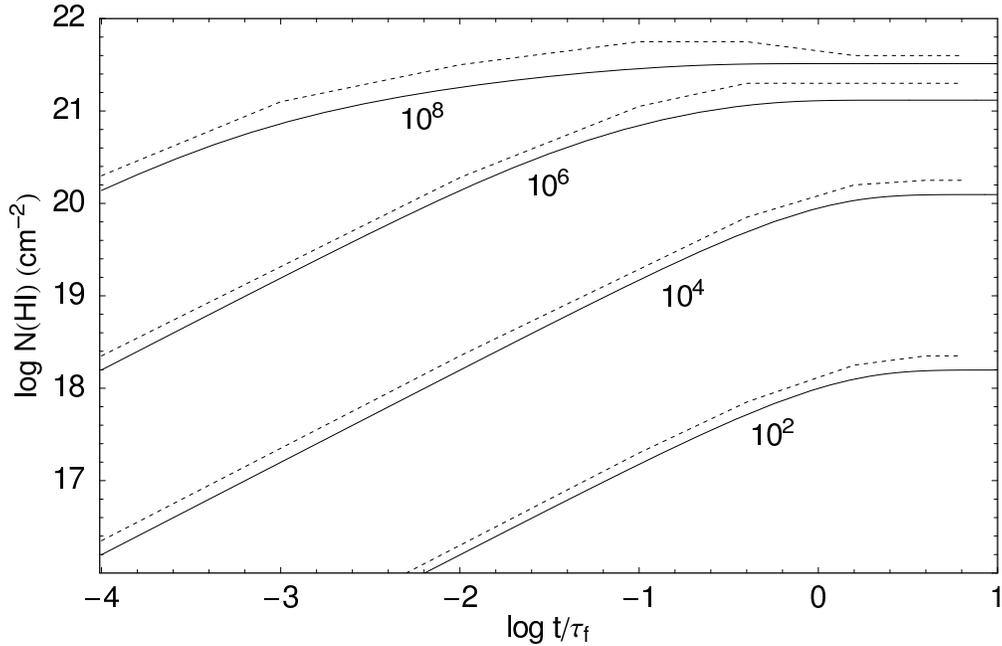}
\caption{Comparison   of   the   solution   of   Eq.    12   for   the
  photodissociation  of  a molecular  (plane-parallel)  cloud and  the
  solution   of   Goldshmidt  \&   Sternberg   (1995)   of  the   full
  integro-differential equations. Plotted is the total HI column N(HI)
  as a function of time since the onset of FUV irradiation (normalized
  on $\tau_{f}$) where  the {\bf drawn lines} are the  solutions  of our
  Eq.   12 for  different  values of the Goldshmidt  \&   Sternberg (1995) $\alpha$ 
  parameter ($\alpha \Phi =\rm r_{dis}$)  while  the {\bf dashed lines} correspond
  to the solutions as read off from their Figure 1.  Lines are
  labelled with  the corresponding  $\alpha$ parameter.  Here  we have
  adopted the same values for $\sigma_d$ and $\Phi$ as Goldshmidt  \&   Sternberg (1995).}
\label{GSfig}
\end{figure}

\newpage

\begin{figure}
\centering
\epsscale{.8}
\plotone{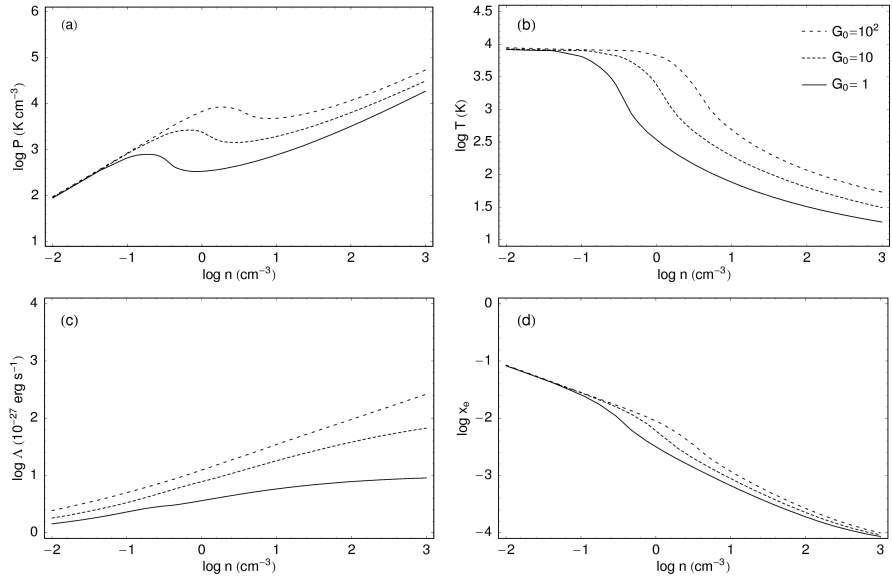}
\caption{Overview of the ISM  model for $\rm Z=Z_{\odot}/5$ (a typical
  metallicity for  dwarf galaxies): equilibrium plots  of (a) pressure
  $\rm P$, (b) temperature $\rm T$, (c) cooling rate $\rm \Lambda$ and
  (d) electron fraction $\rm x_e$ as a function of density $\rm n$ for
  three different values of the UV  field $\rm G_0$ (given in units of
  $1.6 \times 10^{-3}$ erg cm$^{-2}$ s$^{-1}$)}
\label{ismfigm}
\end{figure}
 
\clearpage

\newpage

\begin{figure}
\centering
\epsscale{.45}
\plotone{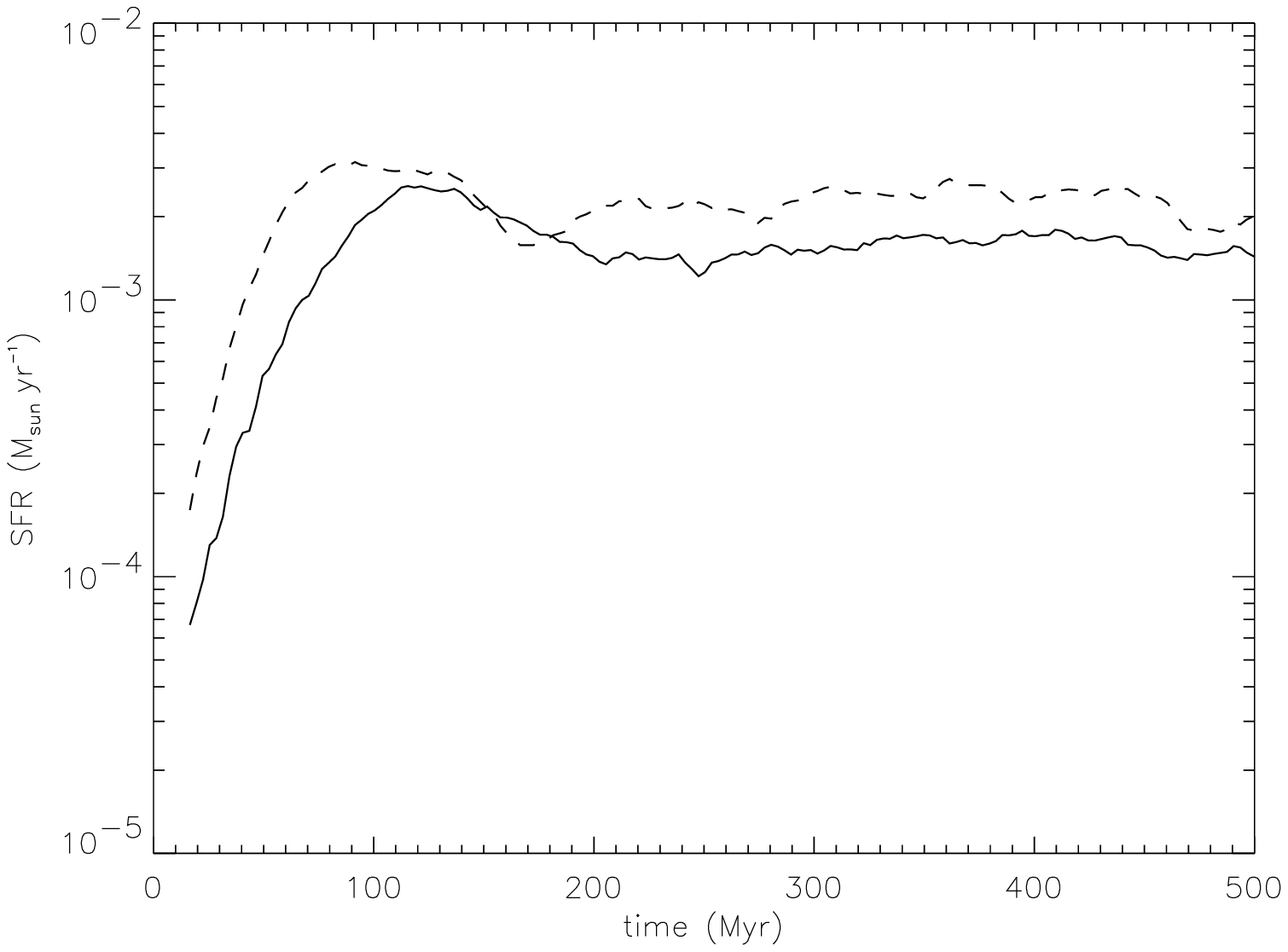}
\plotone{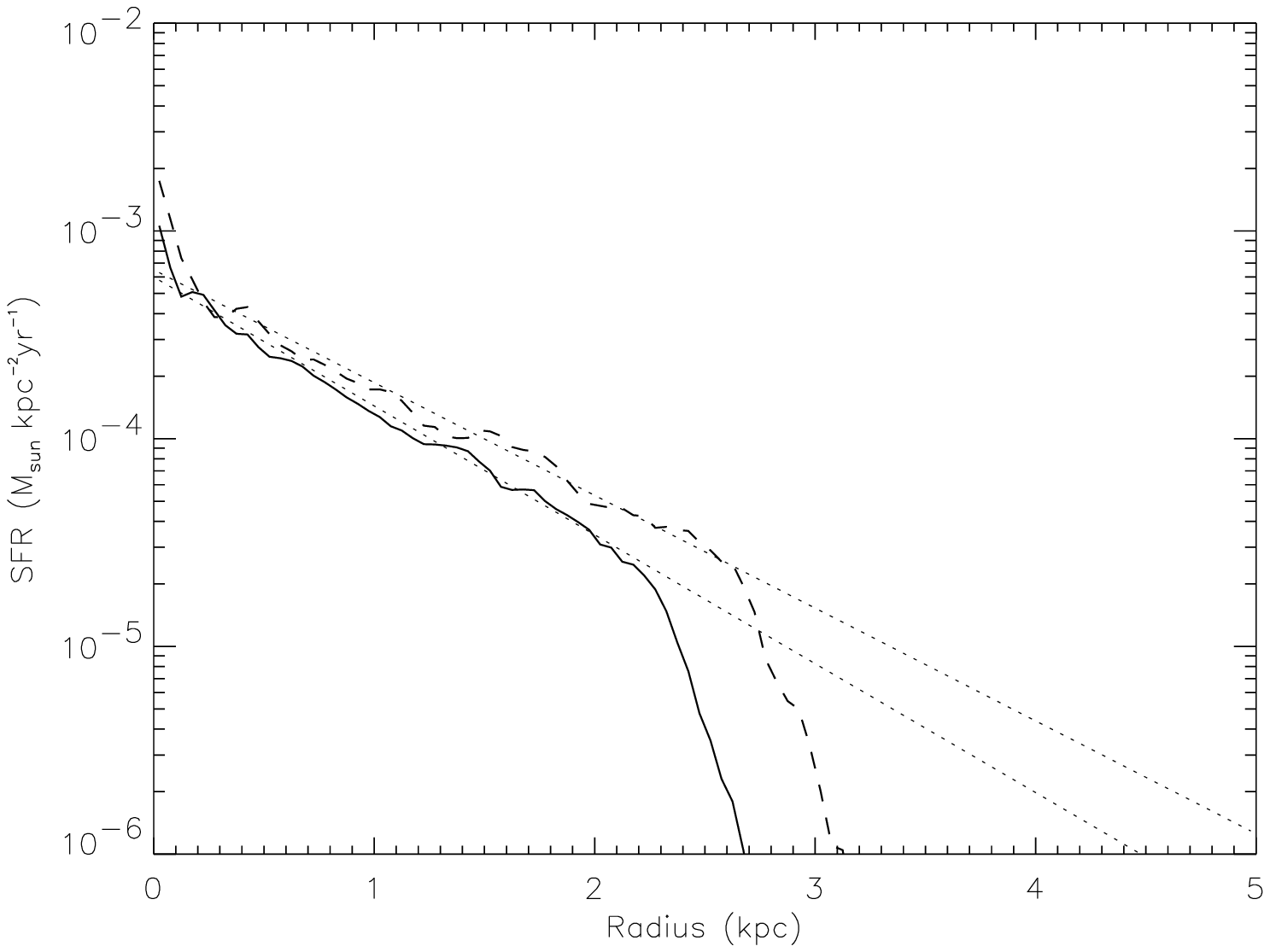}
\caption{Star formation of the model galaxies. Left panel shows the star 
formation as a function of time ({\bf Dotted line}: 
high Z models, {\bf drawn line}: low Z models). The right panel shows  
the mean star formation density as a function of radius (again for both 
models). The dotted lines in the right panels show exponentials with scale 
lengths of 0.7 and 0.75. } 
\label{fg_sf}
\end{figure}

\clearpage

\newpage

\begin{figure}
\centering
\epsscale{.8}
\plotone{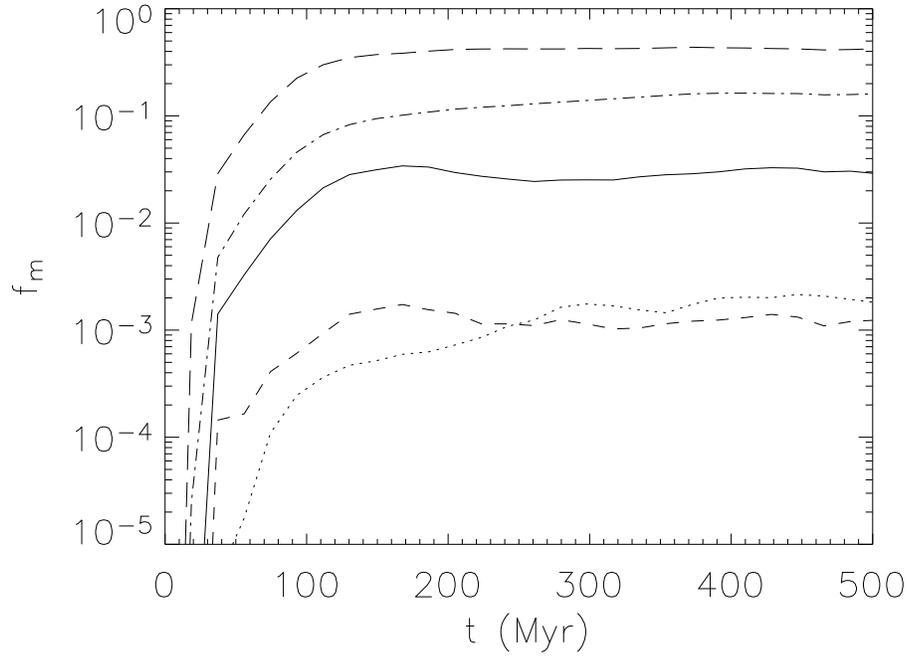}
\caption{H$_2$ fraction vs time. {\bf Dashed line:}  the molecular
fraction $\rm  f_m$ for the $\rm \mu=3.5,  Z=Z_{\odot}/5$ simulation, {\bf
 dash-dotted line:}  $\rm \mu=17.5, Z=Z_{\odot}/5$, {\bf  drawn line:} $\rm
\mu=3.5,  Z=Z_{\odot}$, {\bf long-dashed  line:} $\rm  \mu=17.5, Z=Z_{\odot}$,
and  {\bf  Dotted  line:}  $\rm \mu=3.5,  Z=Z_{\odot}/5$  with  logotropic
clouds.  Simulations are started with $\rm f_m=0$.}
\label{fg_fmtime}
\end{figure}

\clearpage

\newpage

\begin{figure}
\centering
\epsscale{.49}
\plotone{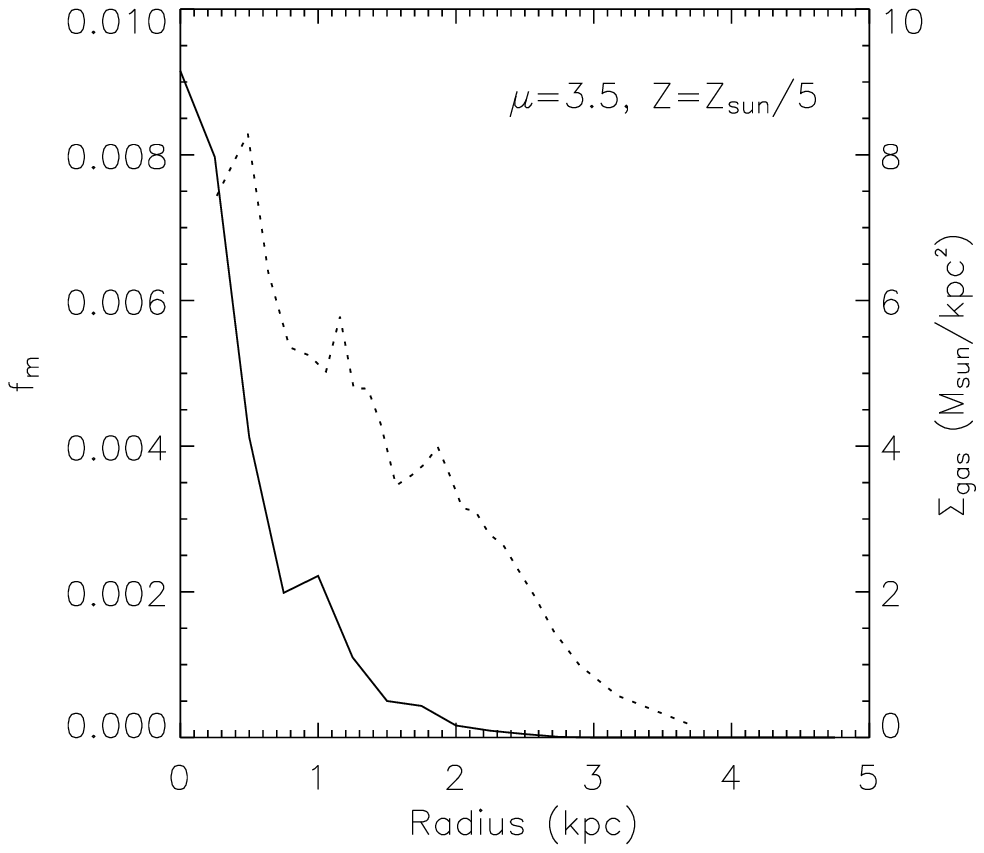}
\plotone{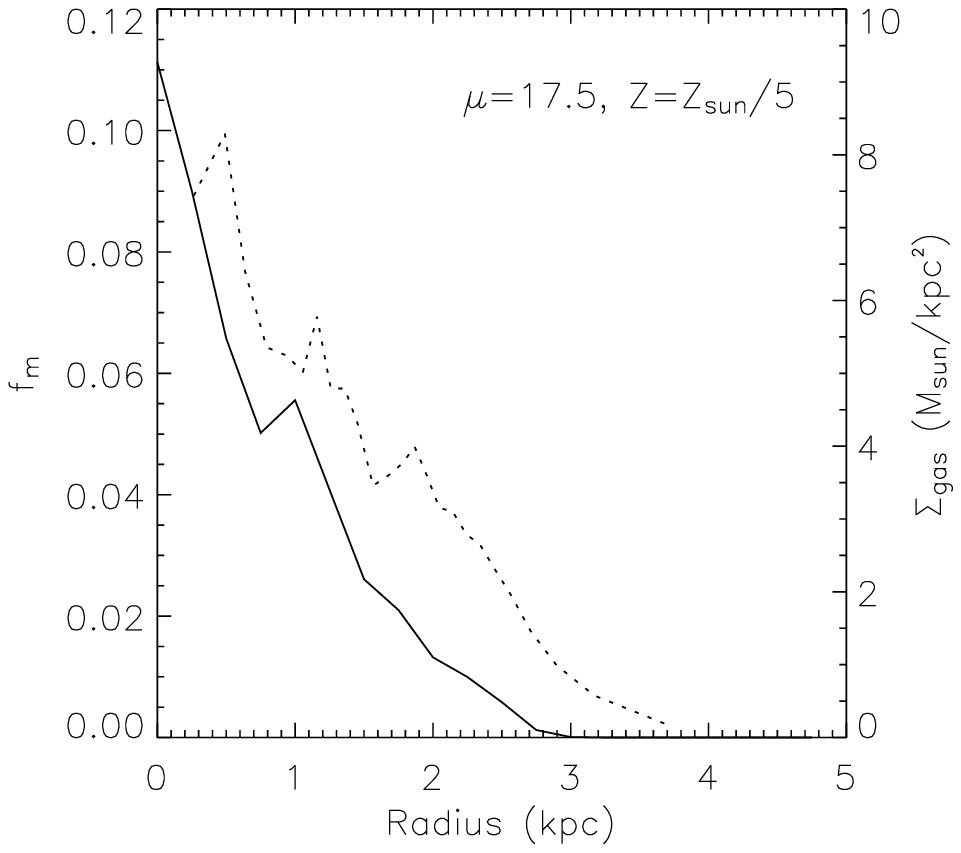}
\plotone{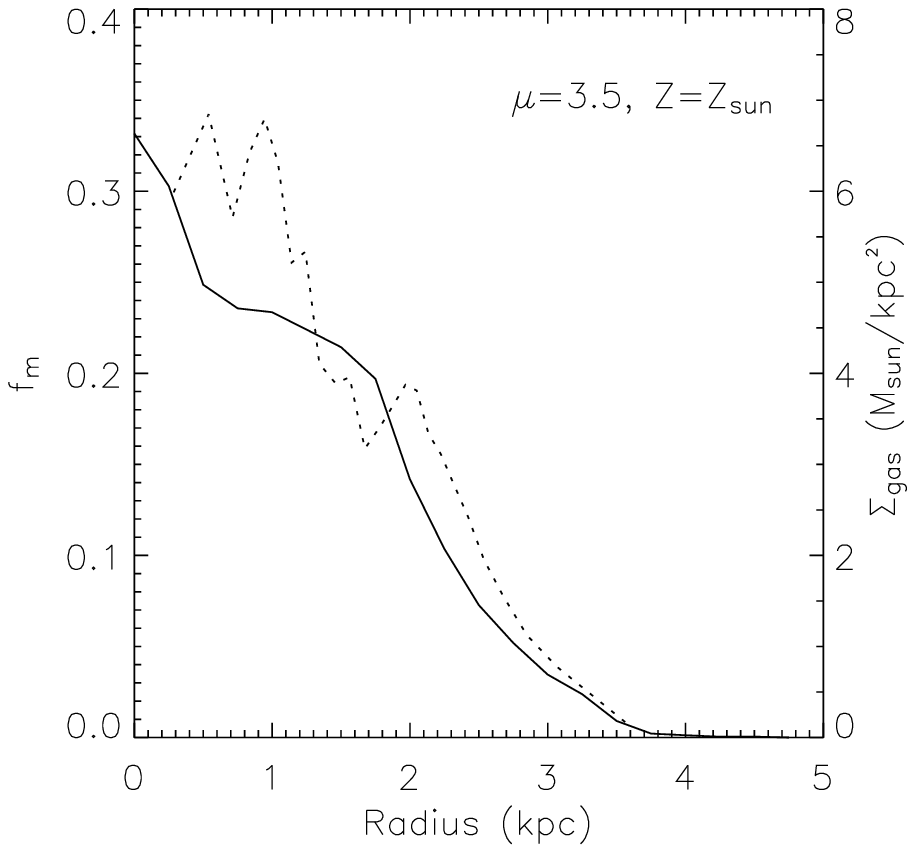}
\plotone{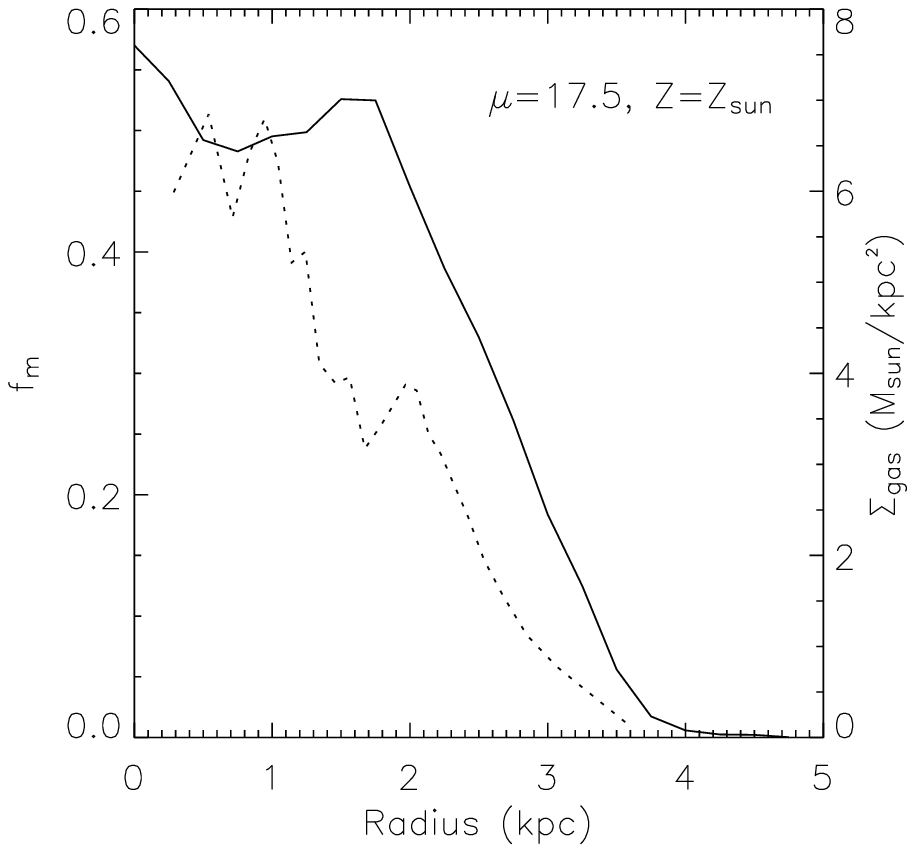}
\caption{Radial dependence  of  the molecular  fraction. {\bf  Drawn
line:} mean molecular fraction as  a function of radius (scale on left
y-axes), and  {\bf dotted line:}  total gas surface density  (scale on
right y-axes).  Note that the y-axis  scales are not the  same for the
different panels.}
\label{fg_rfm}
\end{figure}

\clearpage

\newpage

\begin{figure}
\centering
\epsscale{.49}
\plotone{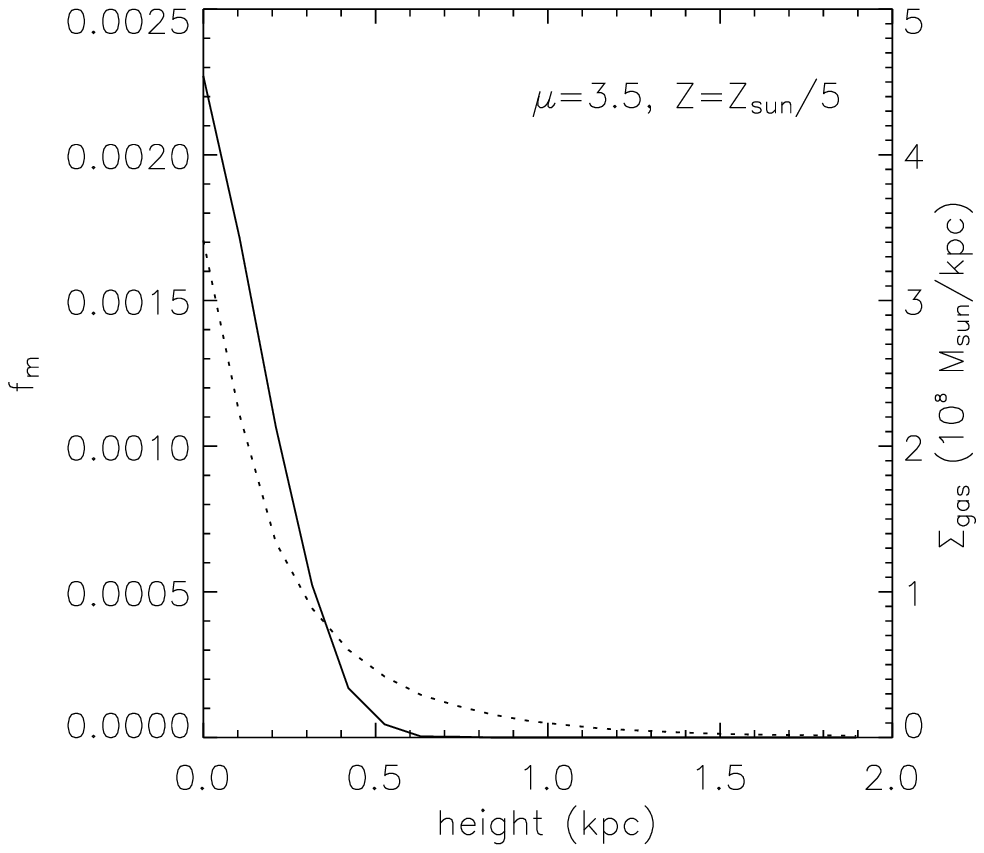}
\plotone{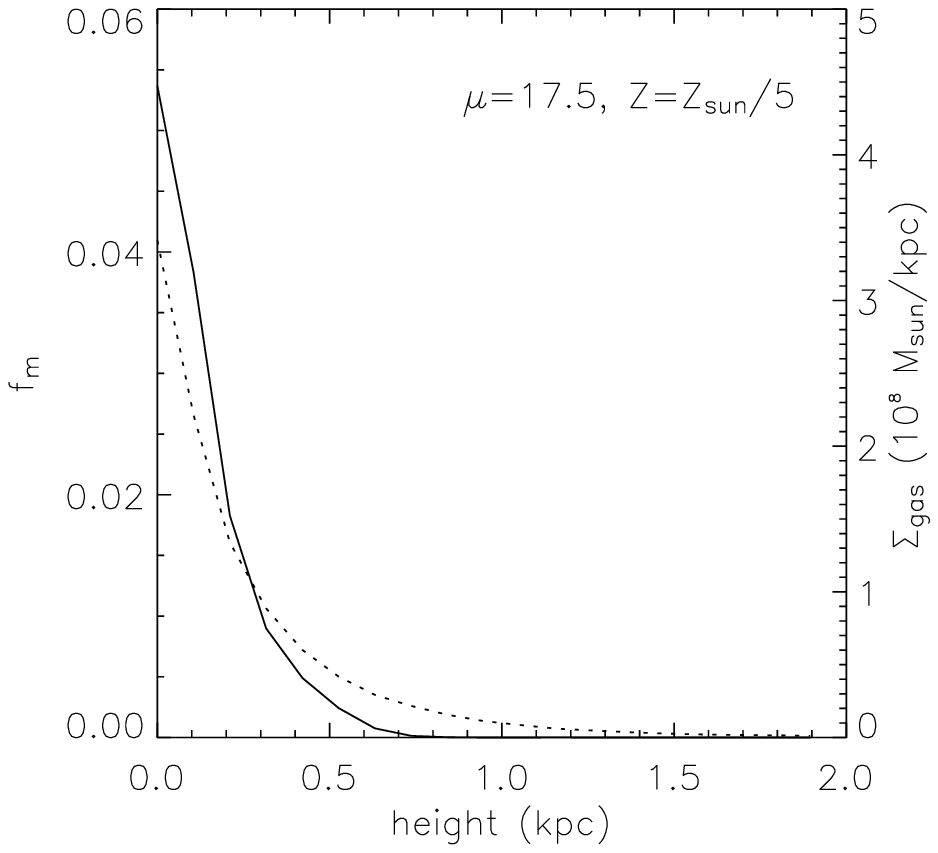}
\plotone{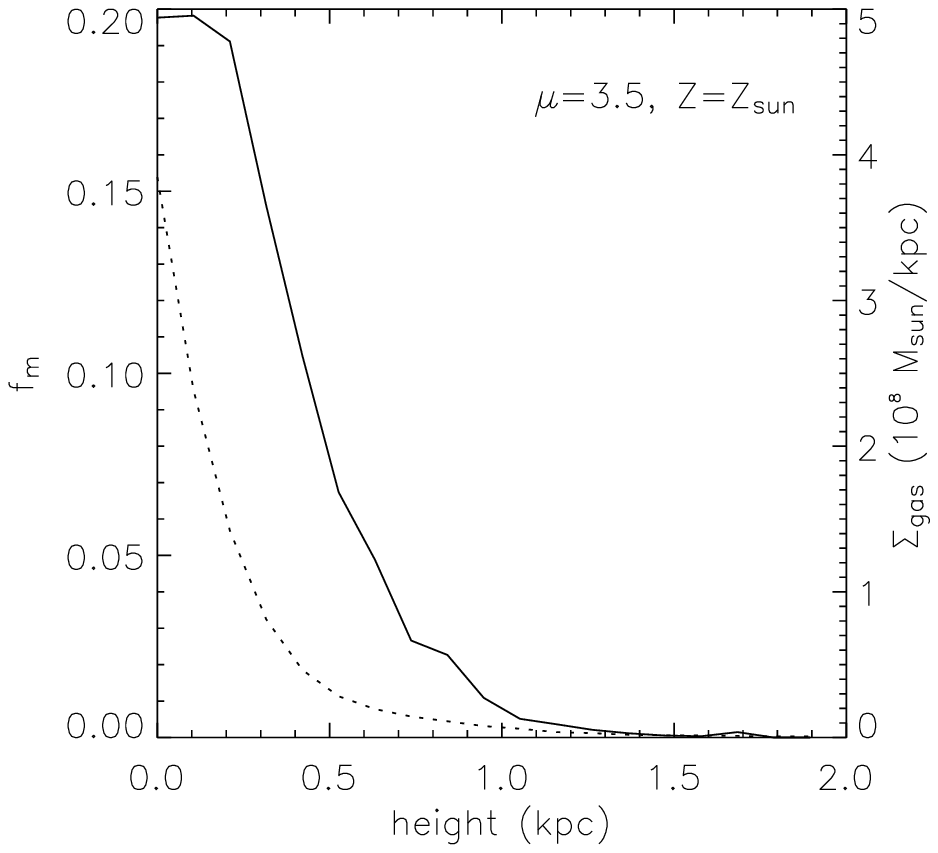}
\plotone{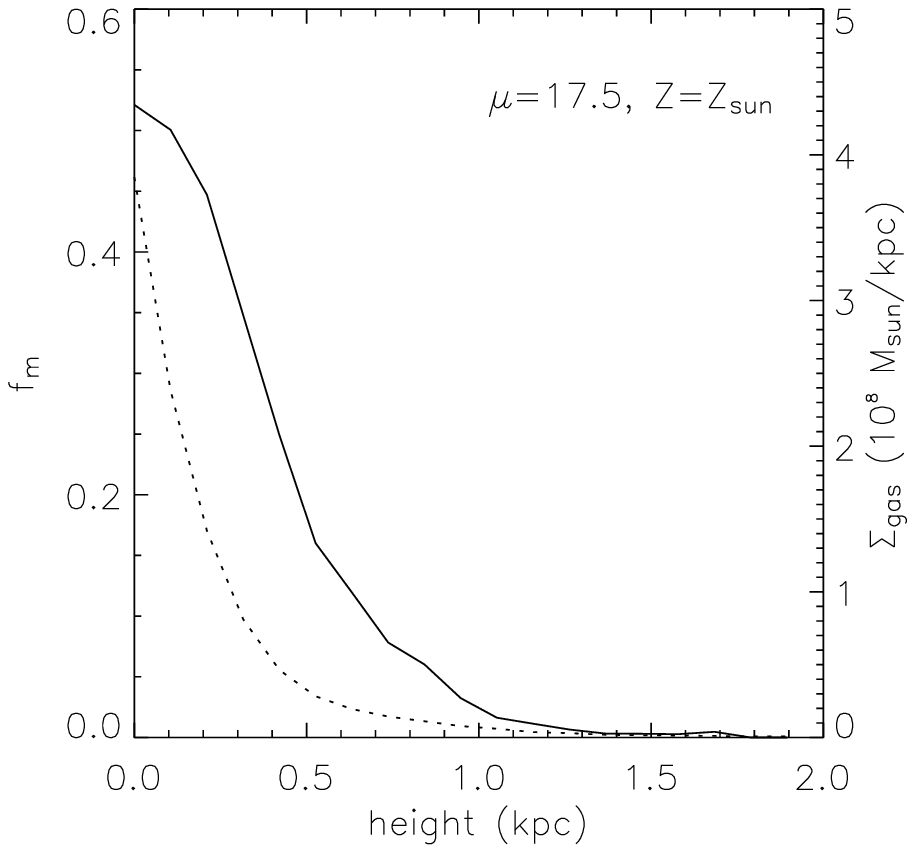}
\caption{z-Height  dependence of  the molecular fraction.  {\bf Drawn
line:} mean molecular fraction as  a function of height above the disk
plane (scale  on left y-axes), and  {\bf dotted line:}  total gas mass
distribution (scale on right y-axes). Note that the y-axis scales are
not the same for the different panels.}
\label{fg_hfm}
\end{figure}

\clearpage

\newpage

\begin{figure}
\centering
\epsscale{.36}
\plotone{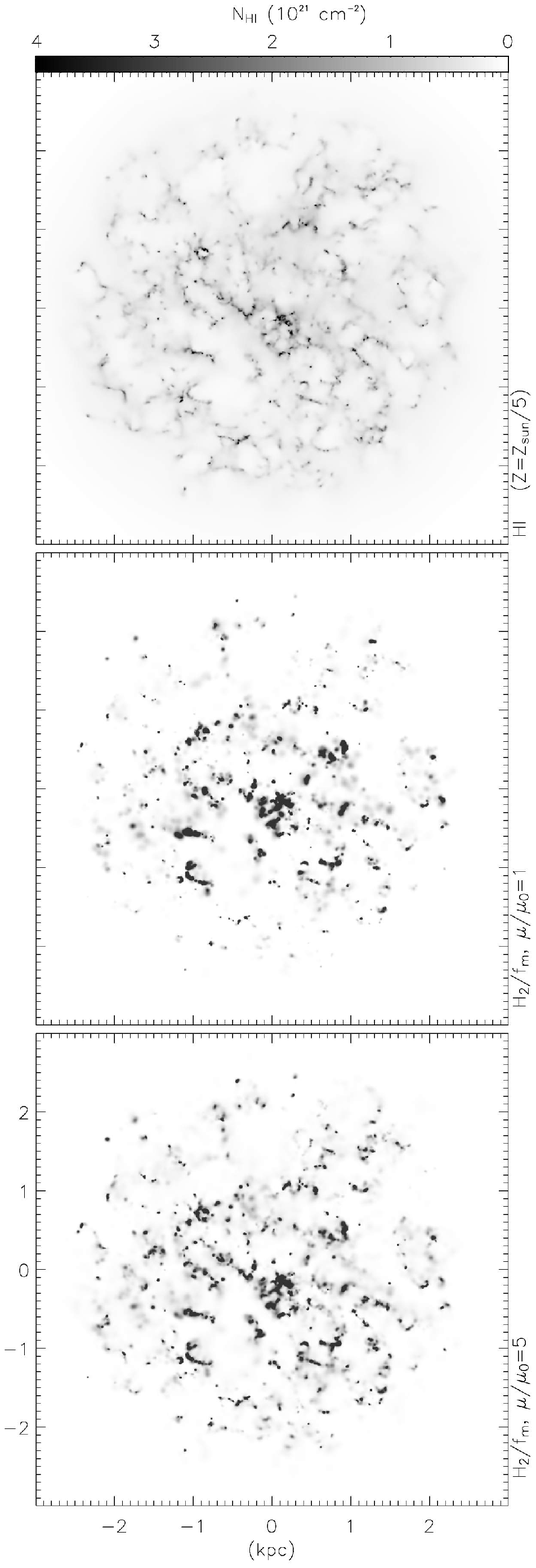}
\plotone{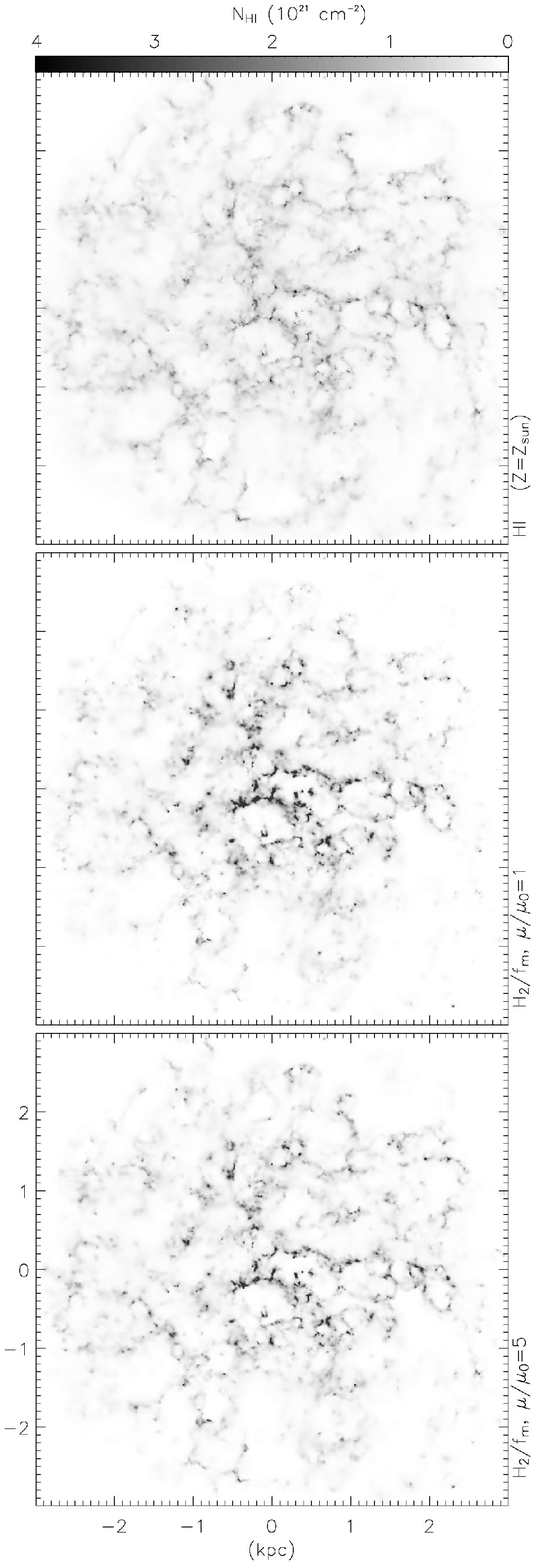}
\caption{HI  and H$_2$  maps.   {\bf Left  panels:}  results for  $\rm
  Z=Z_{\odot}/5$,  {\bf right panels:}  $\rm Z=Z_{\odot}$.  Shown are,
  from top to bottom, the  HI distribution, the H$_2$ distribution for
  $\mu=3.5$ and  for $\mu=17.5$.   The HI maps  of low and  high $\mu$
  simulations are very similar, shown are only the maps for $\mu=3.5$.
  The H$_2$  maps have  been dividided by  the {\it  global} molecular
  fraction such that  the differences with the HI  distribution can be
  seen more clearly. }
\label{fg_hih2maps}
\end{figure}

\clearpage

\newpage

\begin{figure}
\centering
\epsscale{.8}
\plotone{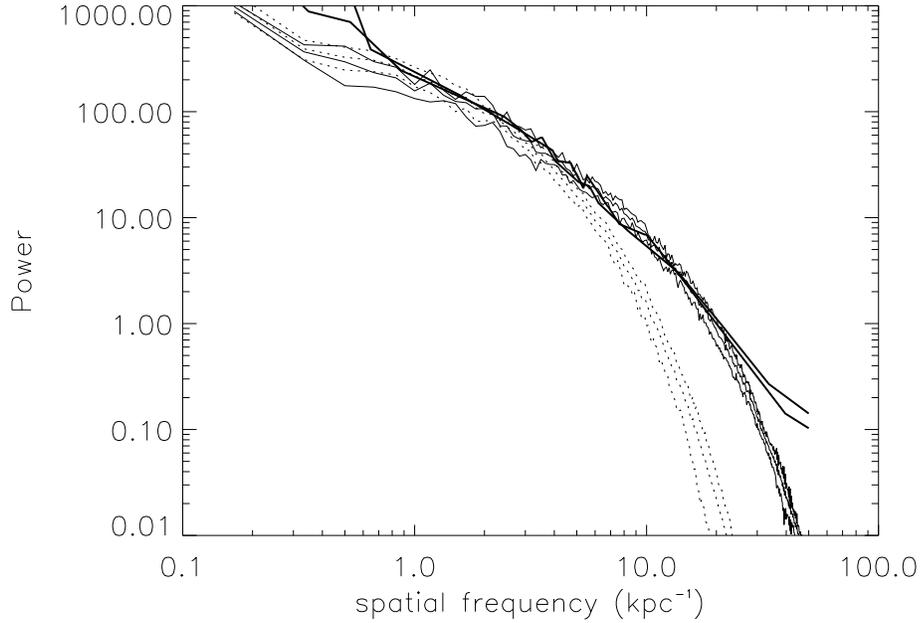}
\caption{ Power spectra of HI maps. {\bf Thick lines} give 1-D 
power spectra of the LMC HI distribution, averaged over all image 
lines in north-south or east-west directions \cite{EKS01}. {\bf Thin 
drawn lines} are the corresponding power spectra made from projected 
HI maps of a simulated dwarf galaxy ($\rm Z=Z_\odot$, $\rm \mu=3.5$).
Given are the mean powerspecrum of 10 snapshots spanning 200 Myr
simulation time and the two lines indicating the mean $\pm 1 \sigma$. 
{\bf Dotted lines} are for the same model at low resolution (N=20k). 
The power is arbitrarily scaled, but the low and high resolution models 
are on the same scale and the LMC spectrum is scaled to match.} 
\label{fg_pw}
\end{figure}

\clearpage

\newpage

\begin{figure}
\centering
\epsscale{.49}
\plotone{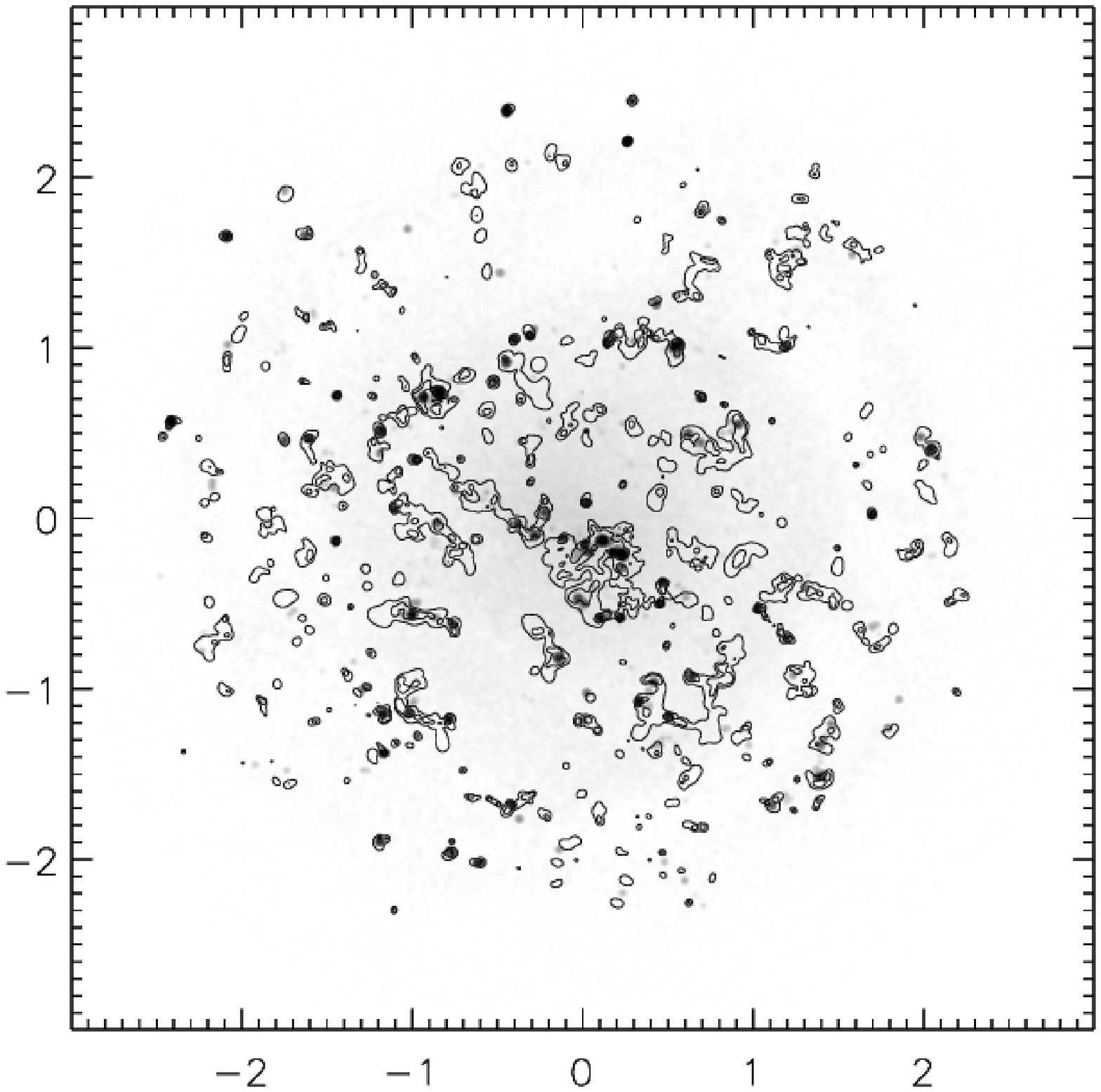}
\plotone{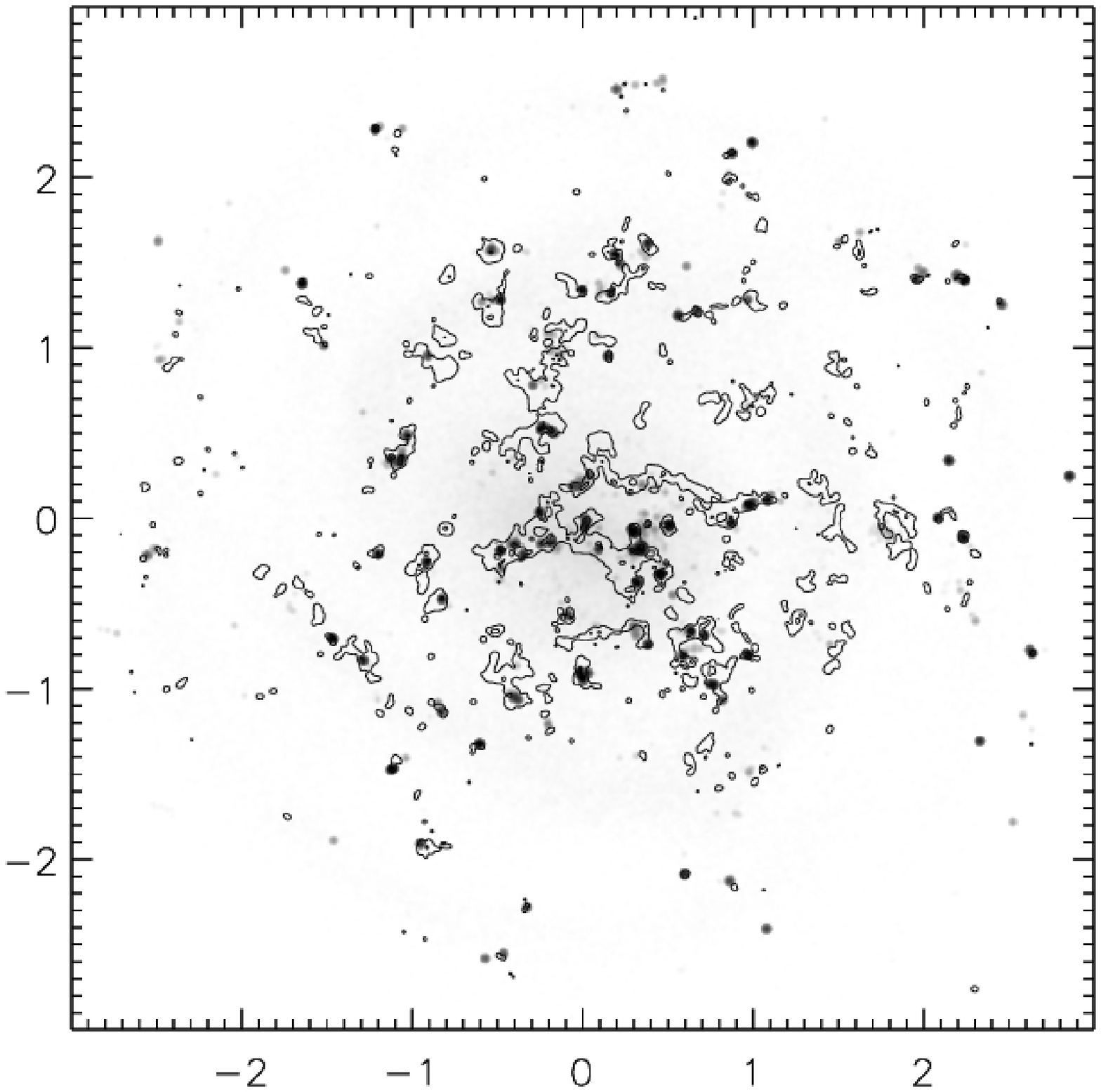}
\caption{H$\alpha$ and  H$_2$: simulated  H$\alpha$ maps  with H$_2$
contours. Left panel shows the  maps for the $\rm Z=Z_{\odot}/5$, $\mu=3.5$
simulation  (contours  at column  densities  of  $\rm N_{HI}=2  \times
10^{19}$  and $10^{20}$ cm$^{-2}$)  while the  right panel  shows the
$\rm Z=Z_{\odot}$,  $\mu=17.5$ run (contours  at column densities  of $\rm
N_{HI}=2 \times 10^{20}$ and $10^{21}$ cm$^{-2}$).}
\label{fg_h2hamaps}
\end{figure}

\clearpage

\newpage

\begin{figure}
\centering
\epsscale{.8}
\plotone{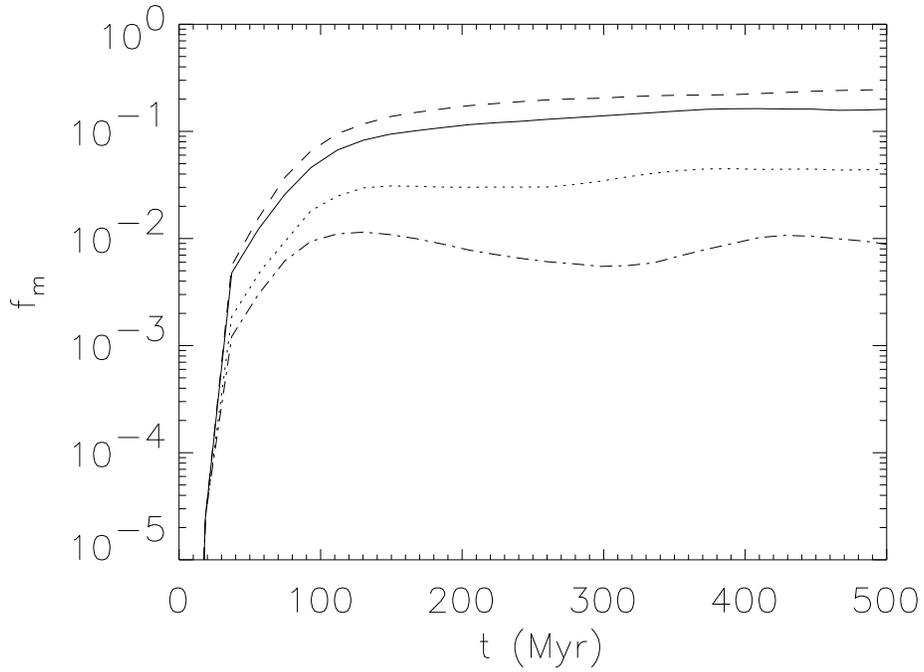}
\caption{H$_2$  fraction vs time for simulations  with different $\rm
f_{sf}$. {\bf  Dotted line:} $\rm f_{sf}=2.5$, {\bf  drawn line:} $\rm
f_{sf}=10$,  {\bf  dashed  line:}  $\rm  f_{sf}=20$. The {\bf dash-dotted} line
gives the molecular fraction for a star formation recipe using a threshold 
$\rm f_{m,sf}=1/8$ ($\mu=3.5$ and $\rm Z=Z_{\odot}$ in all cases).}
\label{fg_fmtime_tcol}
\end{figure}

\clearpage

\newpage

\begin{figure}
\centering
\epsscale{.8}
\plotone{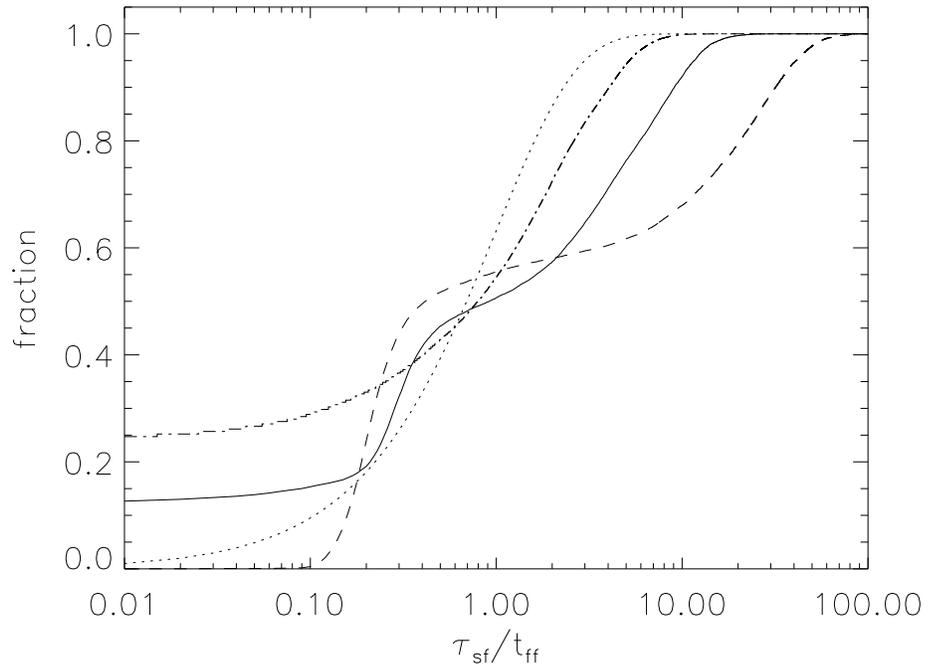}
\caption{Distribution of star formation timescales for molecular 
regulated star formation. Shown are the cumulative distributions 
of star formation delay times, as measured from the time a particle
becomes unstable and normalized on the local free-fall timescale 
$\rm t_{ff}$, for different values of the treshold molecular 
fraction $\rm f_{m,sf}$. {\bf Dash-dotted line:} $\rm f_{m,sf}=1/8$, 
{\bf drawn line:} $\rm f_{m,sf}=3/8$ and {\bf dashed line:} 
$\rm f_{m,sf}=5/8$ (For comparison the dotted line gives an exponential 
distribution).}
\label{fg_tausf}
\end{figure}

\clearpage

\newpage

\begin{figure}
\centering
\epsscale{.8}
\plotone{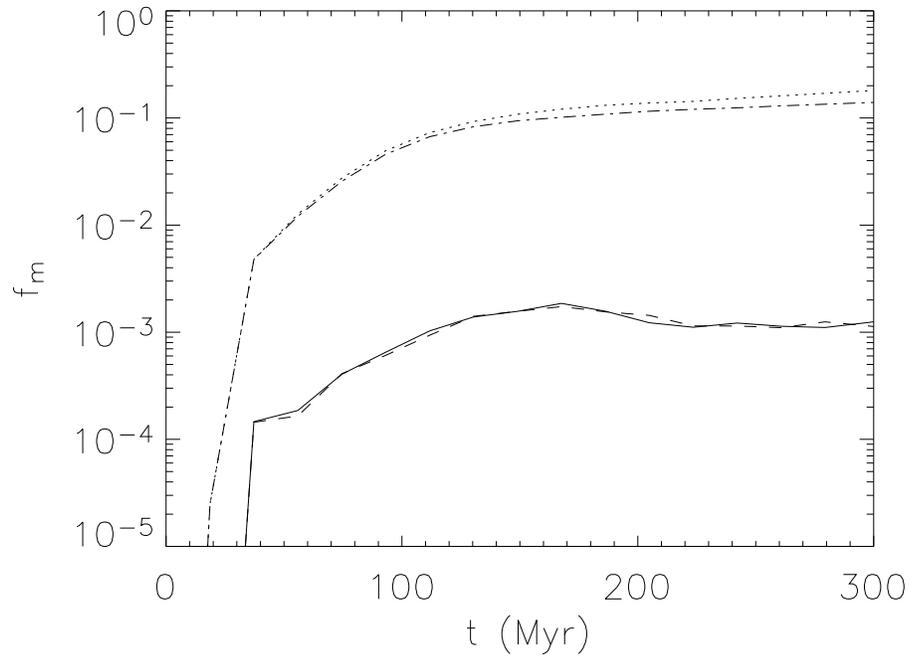}
\caption{H$_2$  fraction vs time  for simulations with  cooling. {\bf
Drawn line:}  $\rm Z=Z_{\odot}/5$, {\bf dashed  line:}$\rm Z=Z_{\odot}/5$ with
H$_2$ cooling, {\bf dash-dotted line:} $\rm Z=Z_{\odot }$ and {\bf dotted:}
$\rm Z=Z_{\odot}$ with H$_2$ cooling. For all: $\mu=3.5$.}
\label{fg_fmcool}
\end{figure}

\clearpage

\newpage

\begin{figure}
\centering
\epsscale{.49}
\plotone{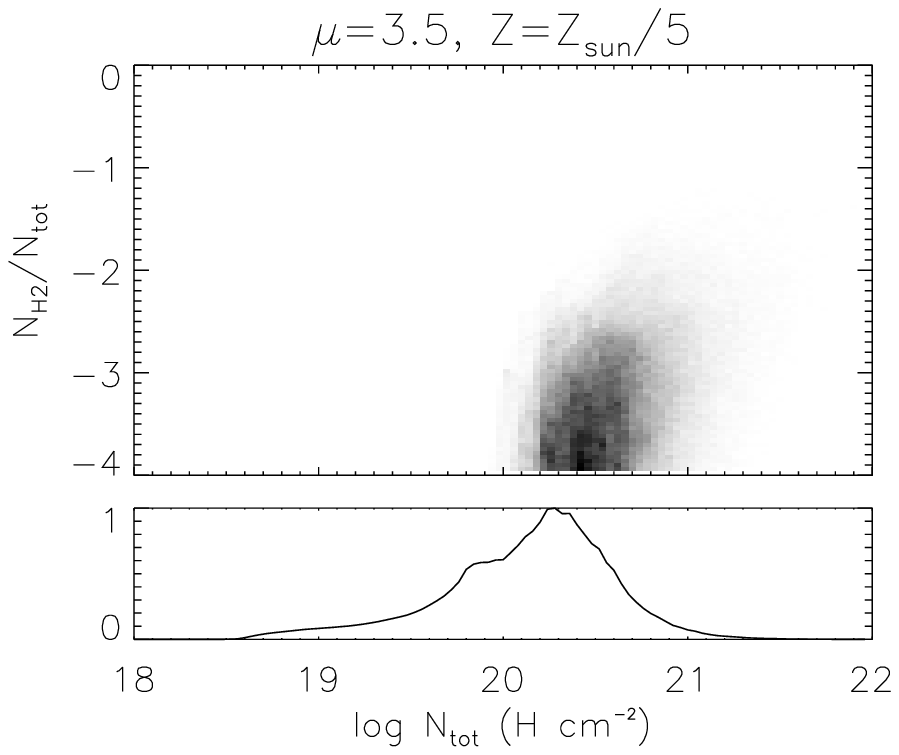}
\plotone{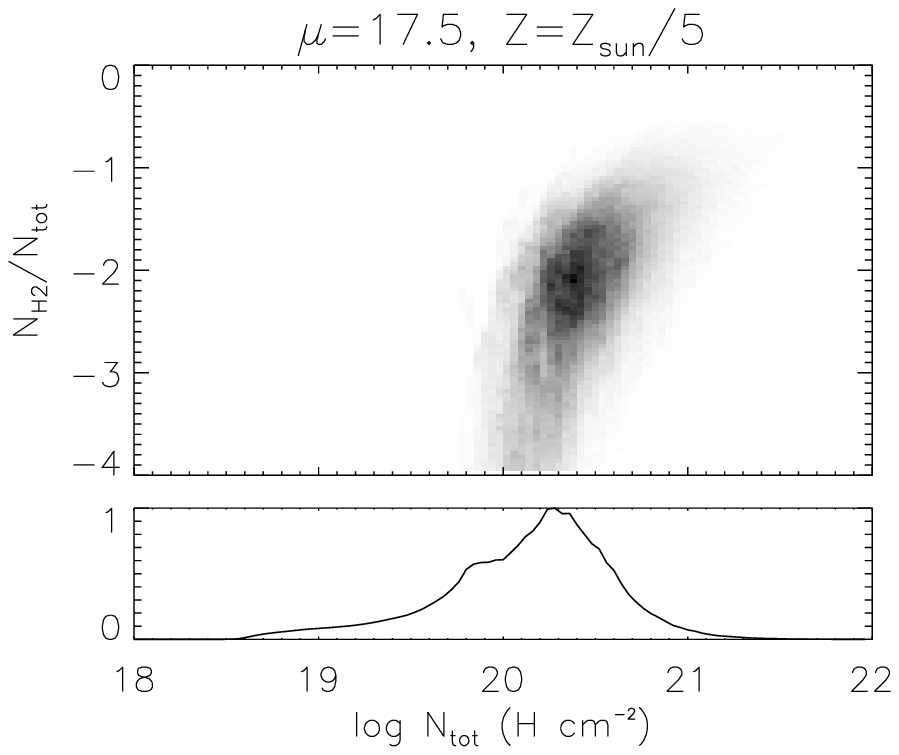}
\plotone{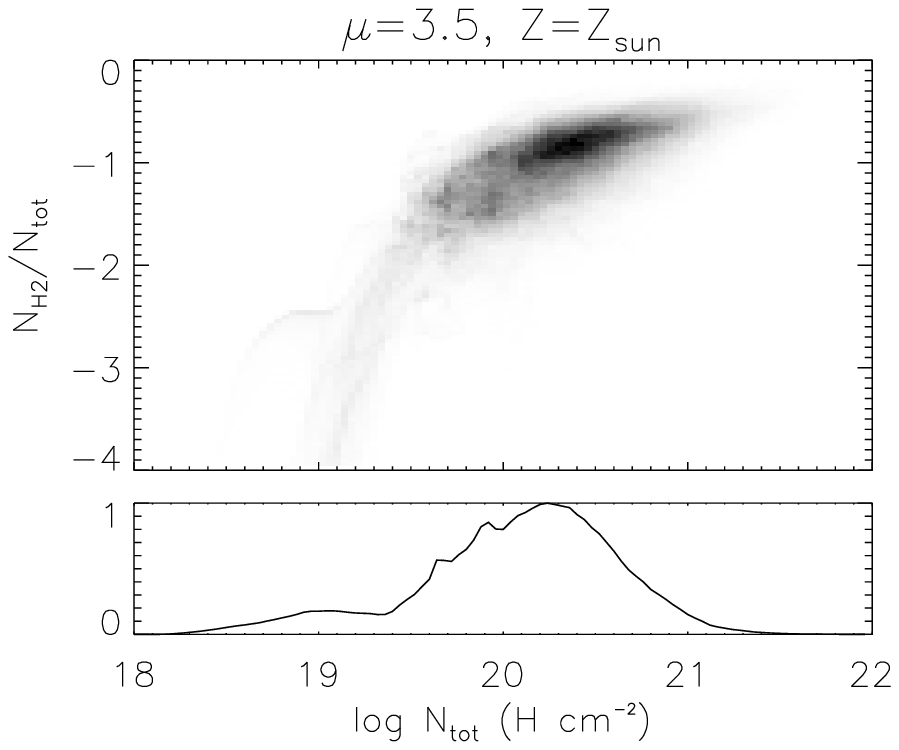}
\plotone{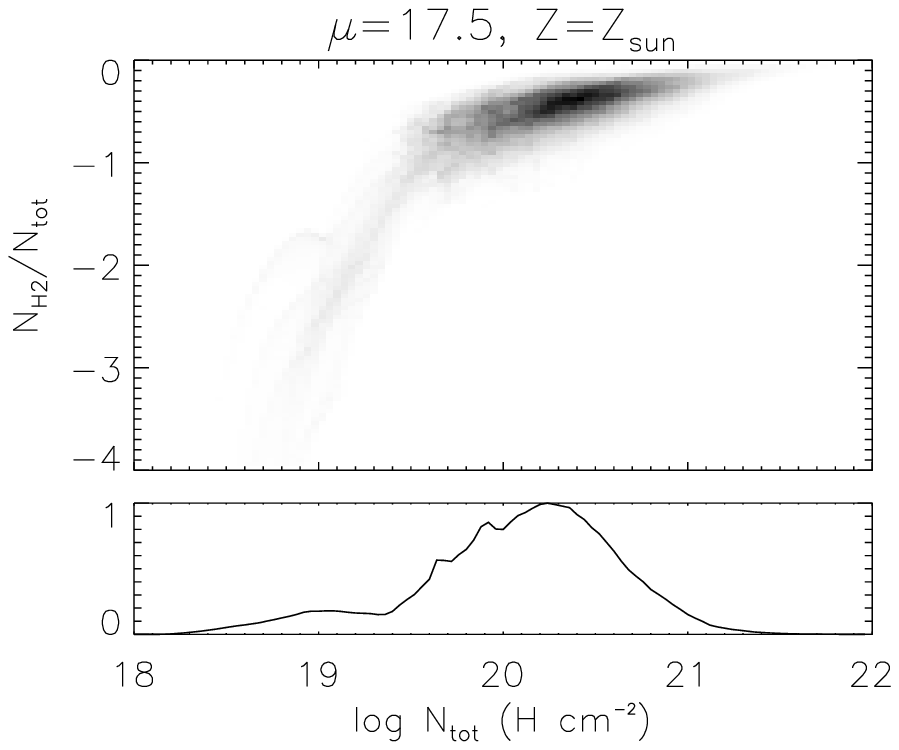}
\caption{ Fraction of H$_2$ vs total column densities. Gray scales give the distribution of H$_2$ fraction $\rm N_{H_2}/N_{HI}$ and total column 
density $\rm N_{HI+H_2}$ of the pixels of HI and H$_2$ projection maps. Histograms give the distribution of pixel values of the total column density, 
scaled to the maximum bin.}
\label{fg_hivsh2}
\end{figure}

\clearpage

\newpage

\begin{table}[t]
\caption[]{  Overview  of the  processes  included  in  the ISM  model
used. For  H and He  ionization equilibrium is  explicitly calculated,
for  other  elements   collisional  ionization  equilibrium  (CIE)  is
assumed.   references: 1: Wolfire et al. (1995), 2: Raga et al. (1997),  3: Verner \&
Ferland (1996), 4: Silva \& Viegas (2001)}

\begin{center}
\begin{tabular}{|l c l|}
\hline
 process &             comment &             ref.    \\
 \hline
 \emph{heating} & & \\
 ~~Cosmic Ray & ionization rate $\rm \zeta_{CR}=1.8~10^{-17}~s^{-1}$   & 1  \\
 ~~Photo Electric & FUV field from stars & 1  \\
\hline
 \emph{cooling} & & \\
 ~~$e$,H$_0$ impact   & H,He,C,N,O,Si,Ne,Fe & 2,4 \\
\hline
 \emph{ionization} & & \\
 \emph{~\& recombination} & & \\
 ~~UV & ionization assumed for & \\
      & species with $\rm E_i< 13.6 \rm{eV}$ &  \\
 ~~Cosmic Ray & H, He only; primary & 1 \\
            & \& secondary ionizations & \\
 ~~Collisional & H, He only & 3 \\
 ~~Radiative recombination & H, He only & 3 \\	    
 ~~CIE & assumed for metals & \\ 
\hline	     	 	 
\end{tabular} 
\end{center}
\label{ism_mtbl}
\end{table}

\end{document}